%% file: hd6e-epjc.tex
\RequirePackage{fix-cm}
\documentclass[twocolumn,epjc3]{svjour3}  
\usepackage{amsfonts}
\smartqed  
\RequirePackage{graphicx}

\RequirePackage{bm}
\RequirePackage{bbm}
\newcommand{\R}{\mathbbm{R}}
\newcommand{\be}{\begin{equation}}
\newcommand{\beq}{\begin{equation}}
\newcommand{\en}{\end{equation}}
\newcommand{\eeq}{\end{equation}}
\newcommand{\bea}{\begin{eqnarray}}
\newcommand{\ena}{\end{eqnarray}}
\newcommand{\hbo}{\hbox to 1 true cm {\hfill } }
\newcommand{\tr}{\hbox{tr}}
\newcommand{\ds}{\delta _s}
\newcommand{\Det}{\hbox{Det}}

\newcommand{\e}{\mathrm{e}}
\newcommand{\lb}{\langle \kern-.17em \langle} 
\newcommand{\rb}{\rangle \kern-.17em \rangle }

\newcommand{\phm}{\phantom{$-$}}
\newcommand{\dlangle}{\left\langle \kern-.17em \left\langle}
\newcommand{\drangle}{\right\rangle \kern-.17em \right\rangle}
\newcommand{\gchi}{$\gg 10^3$}

\journalname{Eur. Phys. J. C}
\begin{document}

\title{Anatomy of the sign-problem in heavy-dense QCD
}


\author{Nicolas Garron\thanksref{e2,addr1} \and 
  Kurt Langfeld\thanksref{e1,addr1,addr2} 
}

\thankstext{e2}{e-mail: nicolas.garron@plymouth.ac.uk}
\thankstext{e1}{e-mail: kurt.langfeld@plymouth.ac.uk}


\institute{Centre for Mathematical Sciences, Plymouth University,
  Plymouth, PL4 8AA, UK \label{addr1} 
  \and
  Theoretical Physics Division, Department of Mathematical
Sciences, University of Liverpool,
Liverpool L69 3BX, UK \label{addr2} 
}

\date{Received: date / Accepted: date}

\maketitle

\begin{abstract}
QCD at finite densities of heavy quarks is investigated using the
density-of-states method. The phase factor expectation value of the
quark determinant is calculated to unprecedented precision as a
function of the chemical potential. Results are validated using those
from a reweighting approach where the latter can produce a significant
signal-to-noise ratio. We confirm the particle-hole symmetry at low
temperatures, find a strong sign problem at intermediate values of the
chemical potential, and an inverse Silver Blaze feature for chemical
potentials close to the onset value: here, the phase quenched theory
underestimates the density of the full theory. 

\keywords{Lattice Gauge theory  \and QCD \and Dense matter  \and Sign problem }

 \PACS{11.15.Ha \and 12.38.Aw \and 12.38.Gc}
\end{abstract}

\section{Introduction}
Monte-Carlo simulations of Quantum Chromo Dynamics (QCD) at finite
baryon densities would provide direct insights into cold, but dense matter as
it occurs in compact stars. They would also trigger the evolution of
effective theories. To date, there are numerous proposals for 
such theories and models. Those rise from exact solvable models
that mimic certain aspects of QCD (see the Gross-Neveu
model~\cite{Schnetz:2004vr,Schnetz:2005ih}) or are motivated by
certain limits of QCD: The limit of many colours has led to the proposal
of the ``quarkyonic
phase''~\cite{McLerran:2007qj,Kojo:2009ha}. Reducing the gluon sector
to the essence of the centre elements has revealed that ``centre-dressed
quarks'' obey Bose statistics and can undergo Bose-Einstein
condensation in the dense, but still confined phase (see
``Fermi-Einstein condensation'' in~\cite{Langfeld:2011rh}). Since
heavy-ion collision experiments probe matter at high temperatures,
but - at best - at moderate densities, the essential input for
understanding cold-dense baryonic matter has to come from first principles
computer simulations. Standard Monte-Carlo simulations,
however, fail since the Gibbs factor is complex at
non-vanishing chemical potentials and, thus, lacks the interpretation
of a probabilistic weight for lattice configurations. This problem
does not exclusively relate to dense QCD, but is generic for dense
matter quantum field theories. It has become known as the notorious 
``sign-problem'' over the last three decades.

\medskip
The recent years, however, have seen significant \break progress in the numerical
studies of complex action systems, both with Monte Carlo methods and techniques that do not rely
on importance sampling. Among the most promising methods are the complexification
of the fields in a Langevin based approach~\cite{Aarts:2011zn,Aarts:2012ft}, 
worm or flux algorithms~\cite{Prokof'ev:2001zz,Alet:2004rh}  to
simulate the dual theory if it happens that this theory is
real~\cite{Endres:2006zh,Endres:2006xu,Mercado:2012ue,Langfeld:2013kno}
and the use of techniques that explicitly exploit the cancellations of classes of
fields~\cite{Chandrasekharan:2010iy}.  

\medskip 
Among the alternatives to conventional Monte Carlo sampling, the
so-called density-of-states simulations (for early results for the
gauge and spin systems see~\cite{Berg:1992qua,Bazavov:2012ex}): this
approach performs Monte-Carlo updates according to the number of
states for a given (complex) action and employs the pioneering
techniques introduced by Wang and Landau~\cite{Wang:2001ab} to refine
the density-of-states during simulation. Once this quantity has been
determined, the partition function and derived expectation values of
observables can be computed semi-analytically by integrating 
the density of states with the appropriate (potentially complex)
Boltzmann weight. More recently, a Wang-Landau type method originally
introduced for continuous systems has been put forward
in~\cite{Langfeld:2012ah,Langfeld:2013xbf,Guagnelli:2012dk}. This
method features an exponential error suppression and allows one to
calculate the density-of-states over many orders of
magnitude~\cite{Langfeld:2015qoa}. At least for the $Z_3$ spin model at
finite densities, the achieved precision of the density-of-states has
been high enough to solve the strong sign problem by direct
integration~\cite{Langfeld:2014nta}. 

\medskip
Heavy-dense QCD (HDQCD) emerges in the limit in which the quark mass and chemical
potential are simultaneously
large~\cite{Bender:1992gn,Blum:1995cb}. This theory has a non-trivial
phase diagram in the plane of temperature and chemical potential, which
qualitatively agrees with the one expected for real QCD: e.g., at
vanishing chemical potential, there  is  a  thermal deconfinement
transition  as  the  temperature is increased with the transition
being first order for very heavy quarks and a crossover for slightly
lighter but still heavy quarks~\cite{Aarts:2014fsa}. The gluonic
part of HDQCD is given by the SU(3) Yang-Mills theory, and a
dualisation that could leave us with a real theory at presence of a 
chemical potentials is not known. So far, this rules out any flux or
worm-type algorithms and makes it a significant testing ground for the
density-of-states techniques. We point out that HDQCD has been
simulated with complex Langevin method providing results for
bench-marking our findings~\cite{Aarts:2014fsa,Aarts:2015yba}. 
We also refer the reader to~\cite{Rindlisbacher:2015pea} 
for a recent study of HDQCD using re-weighting and a mean-field approximation.

\medskip
In this paper, we study HDQCD with the density-of-states approach
detailed in~\cite{Langfeld:2014nta}. The theory is real in the limit
of vanishing and of large chemical potentials and for
chemical potential equalling the heavy quark mass. Although the
phase quenched approximation sketches a qualitatively correct picture for this
reason, we do find a strong sign problem for chemical potentials close
to the mass threshold.

\section{Heavy-dense QCD and the generalised density-of-states approach}

\subsection{HDQCD - definitions and features \label{sec:2.1} }

The partition function of QCD with the quarks field integrated out is
a functional integral over SU(3) unitary matrices only:
\be
Z(\mu) \; = \; \int {\cal D }U_\mu  \; \exp \{ \beta \, S_\mathrm{YM}[U] \}\;
\hbox{Det} M(\mu) \; ,
\label{eq:1}
\en
where we use the Wilson formulation of the Yang-Mills action:
\bea
S_\mathrm{YM}[U] &=& \frac{1}{3} \sum _{x,\mu>\nu} \hbox{Re} \, \tr \Bigl[
  U_\mu (x) \, U_\nu (x+\mu) 
  \nonumber \\
  &&
  \, U^{\dagger} _\mu (x+\nu) \,  U^{\dagger} _\nu (x) \; \Bigr] \; .
\label{eq:2}
\ena
The so-called quark determinant possesses the property
\be
\Bigl(\hbox{Det} M(\mu) \Bigr)^\ast \; = \; \hbox{Det} M(- \mu )
\; , \hbo (\mu \in \R) \; , 
\label{eq:3}
\en
which implies that QCD at vanishing chemical potential, i.e., $\mu
=0$, is a real theory. For large quark mass $m$ and simultaneously
large chemical potential $\mu $, the quark determinant factorises
into~\cite{Bender:1992gn,Blum:1995cb,Aarts:2014fsa,Aarts:2015yba,Rindlisbacher:2015pea}:
\bea
\hbox{Det} \, M(\mu) &=& \prod _{\vec{x}} \; {\det} ^2 \Bigl( 1 \, + \, h
\, \e ^{\mu / T} \, P(\vec{x}) \Bigr)
\nonumber \\
&& {\det } ^2 \Bigl( 1 \, + \, h
\, \e ^{ - \mu / T} \, P^\dagger (\vec{x}) \Bigr) \; ,
\label{eq:4}
\ena
where $T = 1/ N_t a $ is the temperature with $a$ the lattice spacing
and $N_t$ the number of lattice points in temporal direction. The
parameter $h$ is related to the quark hopping parameter $\kappa $ 
and $P(\vec{x})$ is the Polyakov line starting at position $\vec{x}$
and winding around the torus in temporal direction: 
\be 
h \; = \; (2 \kappa )^{N_t} \; , \hbo
P(\vec{x}) \; = \; \prod _{t=1}^{N_t} U_4 (\vec{x},t) \; .
\label{eq:5}
\en
The determinants at the right hand side of (\ref{eq:3}) extend over
colour indices only. Introducing the heavy quark mass $m$ by
\be
m \, a \; = \; - \, \ln (2 \, \kappa ) \; , 
\label{eq:6}
\en
we find that $ h = \exp \{ - m/T \} $ yielding for (\ref{eq:4}):
\bea
\hbox{Det} \, M(\mu) &=& \prod _{\vec{x}} \; {\det }^2 \Bigl( 1 \, + \, 
\, \e ^{(\mu -m) / T} \, P(\vec{x}) \Bigr)
\nonumber \\
&& {\det }^2 \Bigl( 1 \, + \, 
\, \e ^{ - (\mu + m) / T} \, P^\dagger (\vec{x}) \Bigr) \; ,
\label{eq:7}
\ena
Inspection of the latter equation easily confirms that
\be 
\hbox{Det} \, M(\mu=0) \, \in \, \R \; . 
\label{eq:8}
\en
For non-vanishing $\mu $, we will indeed find that the determinant is
complex (albeit the imaginary part can be very small; see
below). However, we are going to show that the partition function is
nevertheless real, i.e., the imaginary part of $Z$ vanishes upon the
integration over gauge configurations. This can be most easily seen by
adopting the Polyakov gauge where
$$
U_4 (t \not=1,\vec{x}) = 1, \hbo P(\vec{x}) \; = \; U_4
(t=1,\vec{x}). 
$$
The partition function takes the form
$$
Z \; = \; \int {\cal D}U_\mu \; \e ^{\beta S_\mathrm{YM}} \; f \Bigl(
U_4(1,\vec{x}), U^\dagger_4(1,\vec{x}) \Bigr) \; , 
$$
where $f$ is a real and analytic function. 
Given that the Haar measure and the action are real, we find upon the
substitution $U_4(1,\vec{x}) \to  U^\dagger_4(1,\vec{x})$ that
\be 
Z (\mu) \; = \; Z^\ast(\mu) \; . 
\label{eq:real}
\en 

For positive chemical potentials and for low
temperatures, i.e.,
\be
\mu \; \ge \; 0 \;, \hbo \frac{m}{T} \gg 1 \; ,
\label{eq:9}
\en
we can neglect quark excitations from the Dirac sea. Formally, 
the second determinant in (\ref{eq:7}) equals unity to a very good
approximation, and we find: 
\bea
\hbox{Det} M (\mu) & \approx & \prod _{\vec{x}} \; {\det }^2 \Bigl( 1 \, + \, 
\, \e ^{(\mu -m) / T} \, P(\vec{x}) \Bigr)
\label{eq:10}
\ena 
For any unitary matrix $P \in SU(3)$, we find that
\be
\det (1 + c \, P) \; = \; 1 + c \, \tr P + c^2 \, \tr P^\dagger + c^3
\; . 
\label{eq:10a}
\en
This implies that the quark determinant is also real for $\mu = m$
(i.e., $c=1$) (see also~\cite{Rindlisbacher:2015pea}): 
\be 
\hbox{Det} \, M (\mu=m) \; \in \; \R \; .
\label{eq:10b}
\en
Let us now study the case of large chemical potentials, i.e., $\mu \gg
m $. Starting from (\ref{eq:10}), we obtain 
\bea 
\hbox{Det} && M (\mu) \; = \; \e ^{2N_c \, V \, (\mu -m) / T}
\label{eq:10c} \\
&& \prod_{\vec{x}} \; {\det }^2 \Bigl( 1 \, + \,  \, \e ^{- (\mu -m) / T} \,
P^\dagger (\vec{x}) \Bigr) \; , 
\nonumber 
\ena
where $N_c=3$ is the number of colours, $V=\sum _{\vec{x}}$ is the
spatial volume  and where we have used that $P$
is a unitary matrix, i.e., $P P^\dagger =1$, $\det P =1$.
It is convenient to introduce the scaled chemical potential relative
to the mass threshold:
\be
t \; = \; \frac{ \mu - m }{T} \; .
\label{eq:11}
\en 
Using (\ref{eq:10}) in the functional integral (\ref{eq:1}), 
the partition function only depends on $t$ and obeys the relation: 
\be
Z (t) \; \approx \; \e ^{2N_c \, V \, t } \; Z (-t) \hbo 
(m\gg T) \; ,  
\label{eq:12}
\en
where we have used that $Z$ is real (see (\ref{eq:real})). As usual, we
define the baryon density by 
\be
\sigma (t) \; = \; \frac{T}{V} \, \frac{ \partial \, \ln \, Z(\mu)
}{\partial \mu } \; = \; \frac{1}{V} \, \frac{ \partial \, \ln \, Z(t)
}{\partial t } \; .
\label{eq:15}
\en
Using (\ref{eq:12}), we find the duality
\be
\sigma (t) \; \approx \; 2N_c \; - \; \sigma (-t) \hbo 
(m\gg T) \; .  
\label{eq:16}
\en
For negative $t$, the chemical potential is below the mass threshold
and the density $\sigma (t)$ rapidly approaches zero with decreasing
$t$. This implies with the help of (\ref{eq:16}) that for large $t$,
the density rapidly approaches the saturation density:
\be
\sigma (t) \stackrel{t \rightarrow \infty} {\rightarrow}\; 2N_c \; .
\label{eq:17}
\en
As a side-remark, we point out that in this regime, i.e., $\mu \gg m$,
the quark determinant becomes a (real) constant (see (\ref{eq:10c})),
$$
\hbox{Det} M (\mu) \; \approx \; \e ^{2N_c \, V \, (\mu -m) / T} \;
, 
$$
and the partition function at large $\mu $ is given by that of pure
$SU(3)$ Yang-Mills theory up to a multiplicative constant. 

\subsection{Reweighting simulations}

If the imaginary part of the quark determinant is small, i.e., for
$\mu \approx 0$ or $ \mu \approx m$ or $\mu \gg m $, the standard
reweighting procedure  can produce reliable results. Using a polar
decomposition of the determinant, the partition function (\ref{eq:1})
can be rewritten as
\be 
Z(\mu)  =  \int {\cal D }U_\mu  \, \e^{ \beta \, S_\mathrm{YM}[U] }\,
\Bigl\vert \hbox{Det} M(\mu) \Bigr\vert \, \exp \{ i \phi [U] \} \, . 
\label{eq:20}
\en
We here introduce the partition function of the phase quenched theory
by 
\be 
Z_\mathrm{PQ}(\mu)  =  \int {\cal D }U_\mu  \, \e^{ \beta \, S_\mathrm{YM}[U] }\,
\Bigl\vert \hbox{Det} M(\mu) \Bigr\vert \, . 
\label{eq:21}
\en
Sometimes, the phase quenched theory is referred to as QCD with an
iso-spin chemical potential. Indeed, rewriting e.g.
\bea 
\Bigl\vert \hbox{Det} M(\mu) \Bigr\vert ^2 &=&
\hbox{Det} M(\mu)  \,  \hbox{Det}^\ast M(\mu)  =
\nonumber \\
&=& \hbox{Det} M(\mu)  \,  \hbox{Det} M(-\mu) \; ,
\nonumber
\ena
the phase quenched theories can be interpreted as (in this case) a
2-flavour quark theory with a chemical potential coupling to the
flavours with opposite sign. 

The Monte-Carlo simulation based upon reweighting generates a Markov
chain of configurations $\{U_\mu\}$ of the phase-quenched theory
(\ref{eq:21}). The expectation value of any observable $A$ is then
obtained by
\be
\langle A \rangle \; = \; \frac{ \langle A \, \exp \{ i \phi [U] \}
  \rangle_\mathrm{PQ}}{\langle \exp \{ i \phi [U] \}
  \rangle_\mathrm{PQ} } \; .  
\label{eq:22}
\en
For a successful reweighting approach, it is essential that the phase
factor expectation value, i.e.,
\be
\Bigl\langle \exp \{ i \phi [U] \} \Bigr\rangle_{PQ} \; = \; \frac{ Z(\mu) }{
  Z_\mathrm{PQ}(\mu) } \; ,
\label{eq:23}
\en
is of significant size. This would ensure a good signal-to-noise
ratio. However, it has been known for a long time (see
e.g.~\cite{Cox:1999nt}), that the full and phase quenched theories
have a difference in their free energy density, say $\Delta f$.
Using the triangle inequality, one also finds that
$$
Z_\mathrm{PQ}(\mu) \; \le \; Z(\mu) \; . 
$$
Hence, the ratio of their partition function in (\ref{eq:23}) is
exponentially suppressed with the volume $V$:
$$
\Bigl\langle \exp \{ i \phi [U] \} \Bigr\rangle_{PQ} \; = \;
\exp \Bigl\{ - \, \Delta f \, V \Bigr\} , \hbo \Delta f \ge 0 \; . 
$$
Consequently, reweighting simulations are restricted to the parameter
space for which the quark determinant is almost real, i.e.,
$$
 \Delta f (\mu) \; \approx \; {\cal O}( 1/V ) \; . 
$$

 \subsection{Density-of-states Method}

 The density-of-states method belongs to the class of Wang-Landau type
 simulations~\cite{Wang:2001ab}. It has been argued
 in~\cite{Langfeld:2015qoa} that the LLR
 version~\cite{Langfeld:2012ah} possesses an {\it exponential} error
 suppression that allows to estimate a strongly suppressed phase
factor expectation value (\ref{eq:23}) with good {\it relative }
precision. This has been demonstrated for the first time for the $Z_3$
spin model at finite densities~\cite{Langfeld:2014nta}.

Central to all Wang-Landau methods is the density of states, which is
defined in the present case of HDQCD by
\be 
\rho (s) \; = \; \int {\cal D}U_\mu \; \delta \Bigl( s - \phi[U]\Bigr)
\; \e^{\beta S_\mathrm{YM}[U] } \; \Bigl\vert \hbox{Det} M \Bigr\vert
\; . 
\label{eq:25}
\en 
Using this definition, the phase factor expectation value
(\ref{eq:23}) can be obtained by Fourier transform
\be
\langle \e^{i \phi} \rangle \; = \; \frac{ \int ds \; \rho (s) \; \exp \{
  i\, s\} }{  \int ds \; \exp \{i\, s\} } \; .
\label{eq:26}
\en
Since the final answer is potentially a very small number, the
density-of-states method needs to overcome two issues here: 
(i) $\rho (s)$ must be calculated to high precision over the whole
range of $s$. This is were standard histogram methods fail: they do
not produce enough statistics in certain regions of $s$ (overlap
problem). (ii) The smallness of $\langle \exp \{i \phi\} \rangle $
arises from cancellations implying that the numerical integration must
be carried out with extreme care. The LLR algorithm generically
overcomes the issue (i), and we refer to the literature for details
(most notably see~\cite{Langfeld:2015fua} for a thorough discussion of
the theoretical framework). To resolve issue (ii), we will adopt
the approach that proved successful in the case of the $Z_3$ spin
model~\cite{Langfeld:2014nta}, and we will present details in the
result section. 

We finally point out that the baryon density $\sigma (\mu)$ can be
calculated once good result for the phase factor expectation value are
available. This rises from the observation that
(\ref{eq:23}) leads to 
\bea
\sigma (\mu) &=&
\frac{T}{V} \, \frac{d}{d\mu } \, \langle \e^{i \phi} \rangle
(\mu) \; + \; \sigma _\mathrm{PQ}(\mu)  \;,
\label{eq:28} 
\ena
where we have introduced the phase quenched baryon density by
\bea 
 \sigma _\mathrm{PQ}(\mu) &=& \frac{T}{V} \frac{ d\, \ln Z_\mathrm{PQ}
   (\mu) }{ d \mu }  \;.
\label{eq:29} 
\ena

\section{Results from reweighting \label{sec:rew} }

\vskip 1mm
\begin{figure*}[t]
\includegraphics[width=7cm]{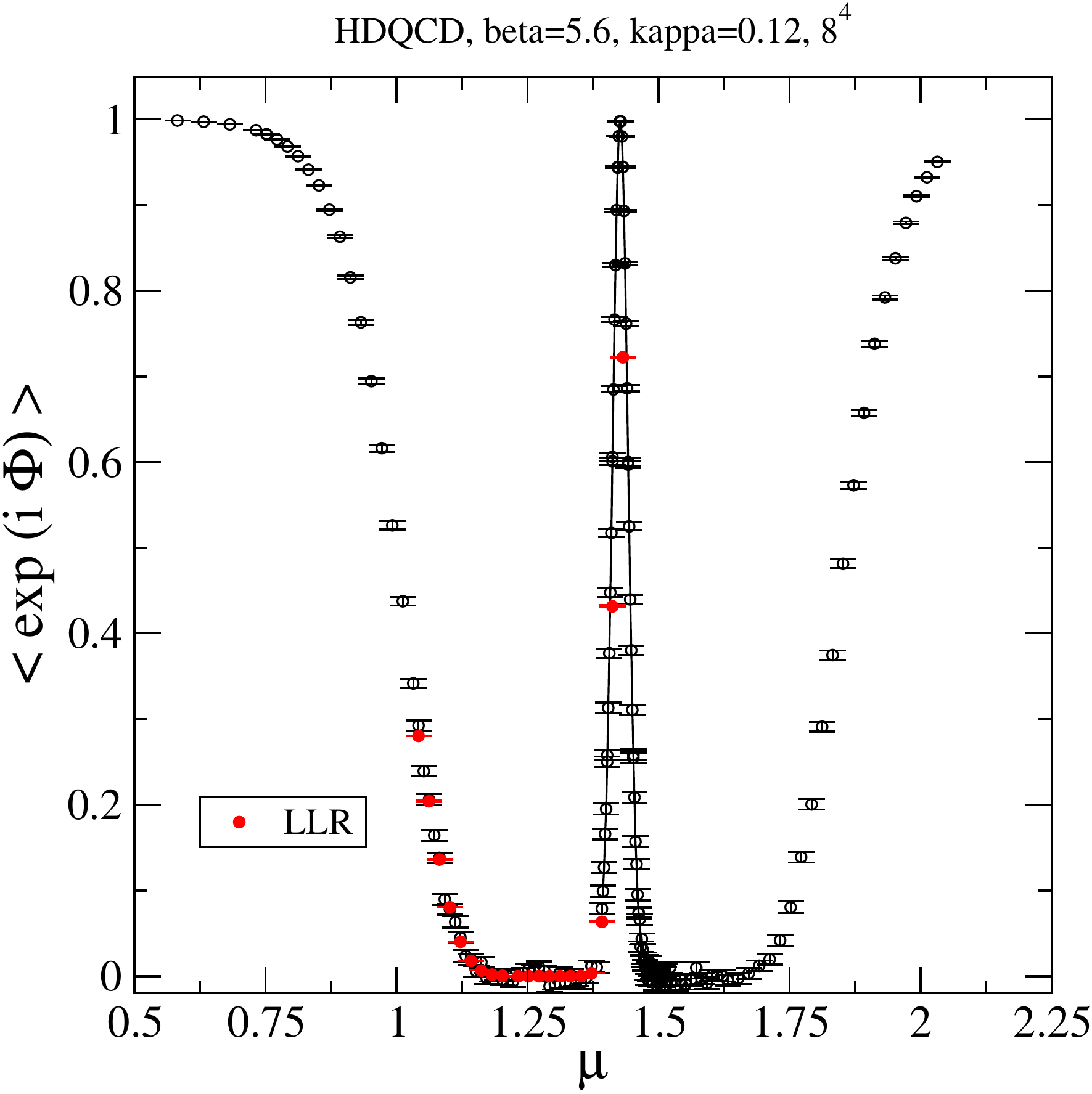} \hspace{0.5cm}
\includegraphics[width=7cm]{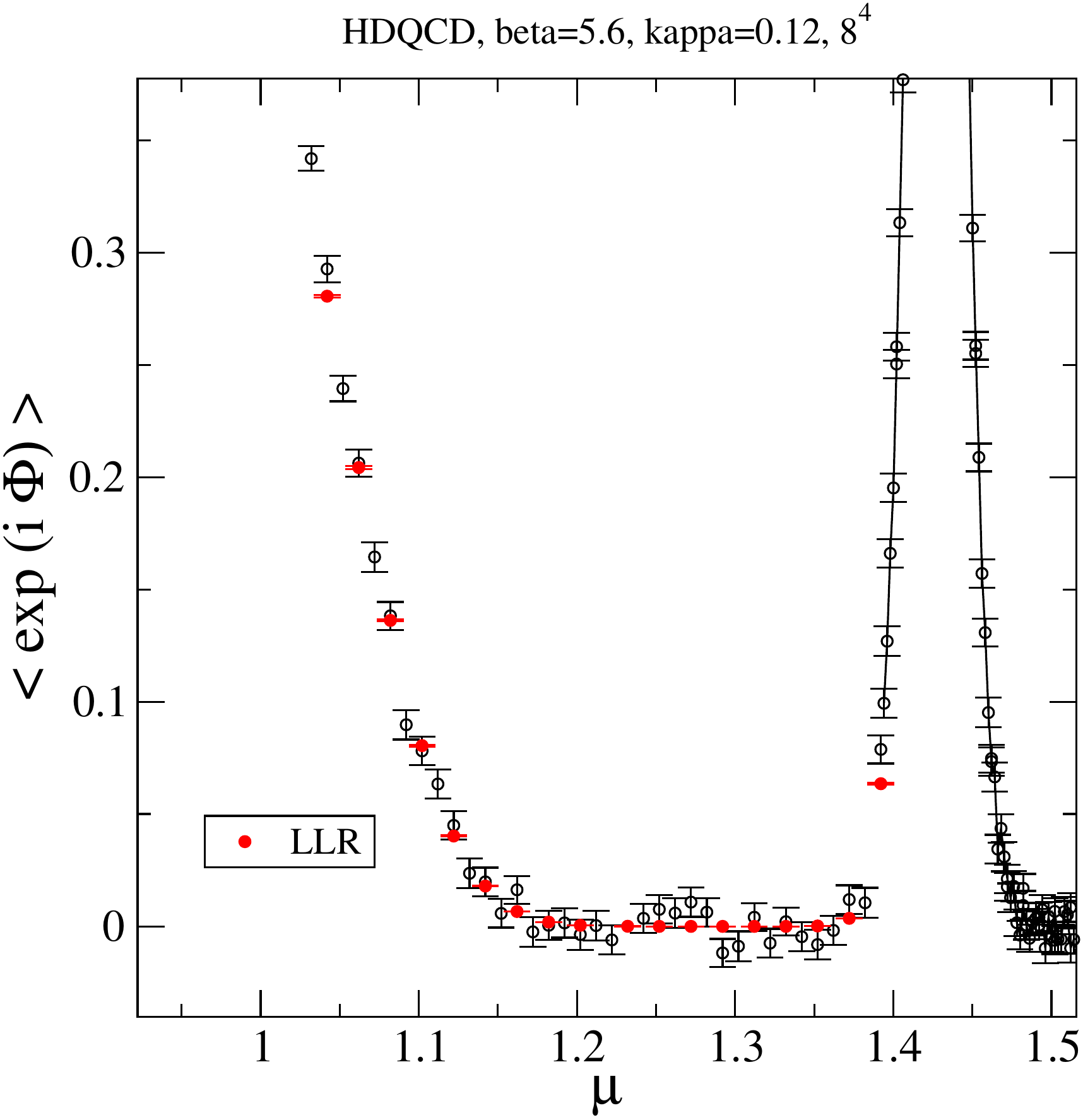} 
\caption{\label{fig:1} Left: The phase factor expectation value $\langle
  \e^{i\phi} \rangle $ as a function of the chemical potential $\mu $
  (simulation parameters in (\ref{eq:31})); Black symbols: the
  reweighting approach; Red symbols: the LLR approach as a
  preview. Right: detail of the graph.   
}
\end{figure*}
Throughout this paper, we use discretised space-time employing a $N^4$
cubic lattice and the Wilson action (\ref{eq:2}).
We work in the Polyakov gauge, i.e., all links are updated except 
$$
U_4(\vec{x},t\not=1) \; = \; 1 \; . 
$$
This implies that the remaining time-like links are identified with
the Polyakov line:
$$
U_4(\vec{x},t=1) \; = \; P(\vec{x}) \; . 
$$
Using the gauge invariance of the quark determinant, it is apparent
that $\hbox{Det} M$ does only depend on $\tr
P^n(\vec{x})$~\cite{Greensite:2013bya,Greensite:2013mfa,Greensite:2014isa}.
We use the Local Hybrid-Monte Carlo (LHMC) simulation algorithm
(with respect to the angles of the algebra) for the update of  
configurations according to the phase quenched partition function
(\ref{eq:21}). We have validated and fine-tuned the algorithm by
comparing some of the results with those obtained by the standard
Cabibbo-Marinari method. The LHMC update shows shorter
auto-correlation times (e.g.~for the topological charge).
The simulation parameters are
\be 
N=8 \; \; \; \; \beta = 5.8 \; \; \; \; \kappa = 0.12 \; \; \; \;
N_\mathrm{conf} = 12 000  
\label{eq:31a}
\en 
where $N_\mathrm{conf}$ is the number of the independent
configurations for the Monte-Carlo estimators. Errors are obtained by
a bootstrap analysis. Our findings from the reweighting approach are
shown in figure~\ref{fig:1}. The chemical potentials are chosen
symmetrically around the mass threshold, which is (using $\kappa = 0.12$,
into (\ref{eq:6}))
$$
a m \; \approx \; 1.427 \;  . 
$$
Our numerical findings are in line with the theoretical predictions
in subsection~\ref{sec:2.1}: the phase factor expectation value
approaches $1$ for small and large values of $\mu $ and for $\mu $
close to the mass threshold. 
Because of the particle-hole duality (\ref{eq:16}), we can confine
ourselves to discussing only the case $\mu \le m$. 
It is remarkable that on a quantitative
level the reweighting approach produces reliable results for $\mu $ as
large as $1$. Note, however, that for the intermediate values, i.e.,
$$
1.15 \; \stackrel{<}{_\sim} \mu \; \stackrel{<}{_\sim} 1.4  \; , 
$$
we do encounter a sign problem with the signal being much smaller than the
noise. 

\begin{figure}
\includegraphics[width=7cm]{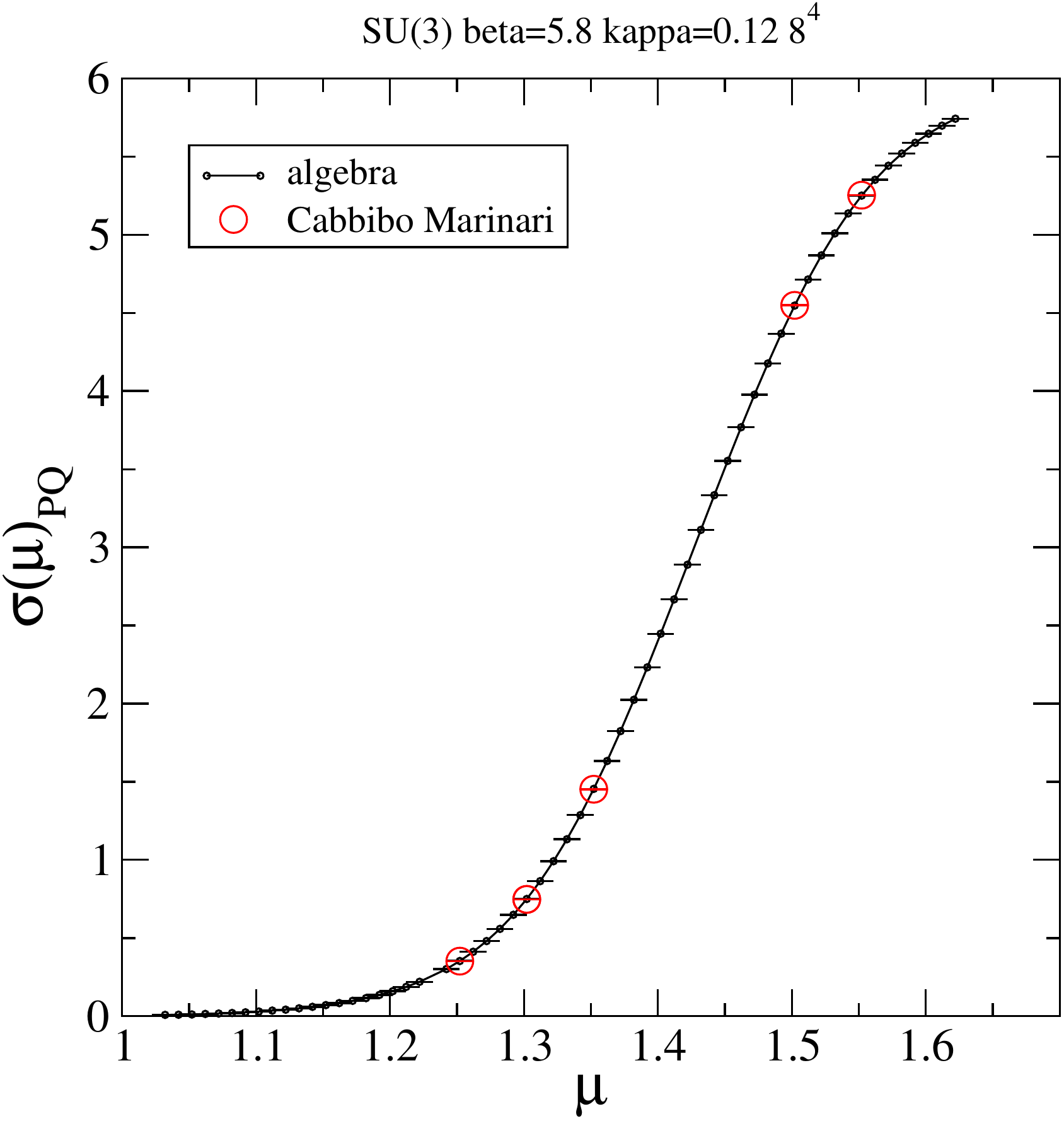} 
\caption{\label{fig:2} Quark density $\sigma _\mathrm{PQ}(\mu)$ of the phase
  quenched theory as a function of the chemical potential $\mu $. 
}
\end{figure}
Let us discuss the implications for the baryon density $\sigma
(\mu)$. We start with a discussion of the phase quenched
density. Since the only $\mu$ dependence is in the quark operator, we
find:
\bea
\sigma _\mathrm{PQ}(\mu) &=&  \frac{1}{Z_\mathrm{PQ}(\mu)}
\int {\cal D }U_\mu  \, \e^{ \beta \, S_\mathrm{YM}[U] }\,
\Bigl\vert \hbox{Det} M(\mu) \Bigr\vert
\label{eq:31} \\
&\times & \frac{\partial }{\partial \mu } \, \ln \, \Bigl\vert \hbox{Det}
M(\mu) \Bigr\vert \; ,
\nonumber 
\ena
where for HDQCD $\hbox{Det} M$ is given in (\ref{eq:4}). As detailed
in subsection~\ref{sec:2.1}, HDQCD is real for vanishing chemical 
potential, for $\mu =m$ and for large $\mu $ implying that
$\sigma = \sigma _\mathrm{PQ}$ for these limiting cases. This signals
that the phase quenched density  shows the correct behaviour for small
$\mu $, the correct onset at $\mu =m$ and the correct asymptotic value
given by saturation. It is therefore expected that $\sigma
_\mathrm{PQ}(\mu) $ qualitatively reflects the $\mu $ dependence of
the full density $\sigma $. This is indeed verified by our direct
evaluation of (\ref{eq:31}) shown in figure~\ref{fig:2}. Although phase
quenching produces qualitatively correct results, we cannot conclude
that the sign problem is weak (see below).

Regardless of the quantitative details, we can draw some interesting
conclusions for the density using the identity (\ref{eq:28}).
For small chemical potentials, e.g., $\mu \le 1.1$, the phase factor
expectation value is decreasing. Consequently, the correction
$$
\frac{T}{V} \, \frac{d}{d\mu } \, \langle \e^{i \phi} \rangle
$$
is negative implying that the phase quenched result overestimates the
true result $\sigma $. This is usually referred to as ``Silver Blaze
problem''. With a smoothness assumption of $\langle \e^{i \phi}
\rangle$, we expect that its derivative with respect to $\mu $
vanishes at $\mu _{c1}$ with $1.15 < \mu _{c1} < 1.4$. For this
chemical potential, we find agreement:
$$
\sigma _\mathrm{PQ} (\mu_{c1}) \; = \; \sigma (\mu_{c1})    \; . 
$$
For $\mu _{c1} < \mu < m$, the derivative of $\langle \e^{i \phi}
\rangle$ is positive. We here find an {\it inverted Silver Blaze
  behaviour}: close to the mass threshold, the phase quenched theory 
       {\it underestimates} the value of the density.

\section{LLR results}

\subsection{Foundations of the  LLR simulation}
\label{s:foundations}
Our aim is to calculate an approximation of the density-of-states
$\rho (s)$ for the imaginary part $s$ of the quark determinant.
We divide the domain of support for $\rho $ into intervals $[s_k,
  s_k+ \ds]$. Under physically motivated assumptions, $\rho (s)$ is a
smooth function such that a Taylor expansion over these intervals
yields a valid approximation. Central to the LLR approach are the
Taylor coefficients (also called LLR coefficients)
\be 
a_k :=  \frac{ 
d \; \ln \, \rho }{ds } \Big\vert _{s=s_k+ \ds/2} \; ,  
\label{eq:40}
\en
which will be the target of our numerical simulations below. With
these coefficients at our fingertips, we use a piece-wise linear
approximation for $\ln \rho $ and derive the approximation: 
\bea 
\rho (s) &=& 
\rho_0 \left( \prod_{k=1}^{N-1}  \e^{a_k \ds} \right) \;
\exp \left\{ a_N  \, \left(s  - s_N \right) \; \right\}  \; , 
\label{eq:41}
\ena
where, for a given $s$, the upper boundary $N$ is chosen such that 
$$ 
s _N \leq s \leq s_N \, + \, \ds \; , \hbo 
s_k \; = \; s_0 \; + \; k \, \ds \; . 
$$
The goal of the LLR method is to calculate the coefficients from a
stochastic non-linear equation. A key ingredient of this equation is 
the restricted and reweighted expectation
values~\cite{Langfeld:2012ah} with $a$ being an external variable
(not to be confused with the lattice spacing): 
\bea
\dlangle W [\phi] \drangle_{k} (a) &=&
\frac{1}{{\cal N}_k} \int {\cal D} U_\mu \; \Big\vert \Det M \Big\vert
\, \e^{\beta   S_{YM} } 
\nonumber \\
&& \theta _{[s_k,\ds]}(\phi[U])\; W[\phi] \; \, \; \; \e^{-a  \phi [U] } \; , 
\label{eq:42}
\ena
\bea 
N_k &=& \int {\cal D} U_\mu \; \Big\vert \Det M \Big\vert \, \e^{\beta
  S_{YM} } \; \theta _{[s_k,\ds]}(\phi[U]) 
\nonumber \\
&& \; \, \; \; \e^{-a   \phi [U] } \; , 
\label{eq:43} 
\ena
where we have introduced the modified Heaviside function 
$$ 
\theta _{[s_k,\ds]} (\phi ) \; = \; \left\{ \begin{array}{ll} 
1 & \hbox{for} \; \; \; s_k \leq \phi  \leq s_k + \ds \\ 
0 & \hbox{otherwise . } \end{array} \right. 
$$
For the particular choice
$$
W[\phi] \; = \; \phi \, - \, s_k \, - \, \frac{\ds}{2} \; =: \; \Delta
\phi 
$$
we showed that 
\be 
\dlangle \Delta \phi \drangle_{k} (a) \; = \; 0 \hbo \hbox{for} \hbo
a \; = \; a_k \; .
\label{eq:44} 
\en
The latter equation is a non-linear equation to obtain $a$. For
instance, this can be done by using the fixed point iteration:
$$
a_k^{(n+1)} \; = \; a_k^{(n)} \; + \; \frac{12}{\ds ^2} \; \dlangle \Delta
\phi \drangle_{k} \left( a^{(n)}_k \right) \; . 
$$
\begin{figure}
\includegraphics[width=7cm]{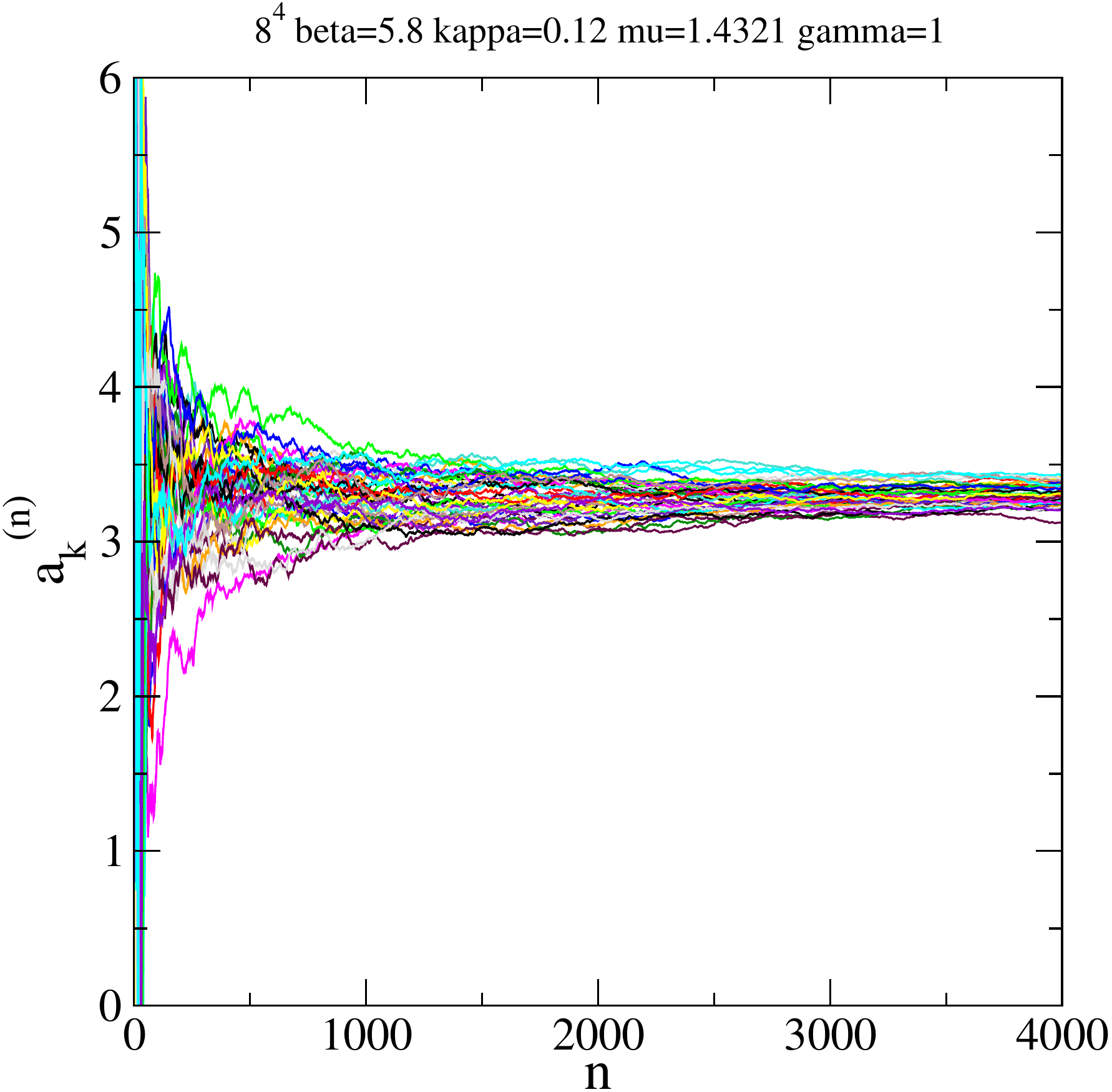} 
\caption{\label{fig:3} Thermalisation history (simulations parameters
  are in table~\ref{tab:1}). 
}
\end{figure}
Note that  the expectation value $\dlangle \Delta \phi \drangle_{k} $
is not known exactly. An estimate, however, can be obtained by
standard Monte-Carlo  simulations. The issue here is that the
statistical error interferes with convergence of the fixed point
iteration. The mathematical framework to obtain a solution was
developed by Robbins and Monro. They showed that the {\it under
  relaxed} iteration
\bea
a_k^{(n+1)} &=& a_k^{(n)} \; + \; \alpha _n \, \frac{12}{\ds ^2}
\; \dlangle \Delta \phi \drangle_{k} \left( a^{(n)}_k \right)
\label{eq:45} \\
\sum_n \alpha _n &\to & \infty \; , \hbo 
\sum_n \alpha ^2_n  \; = \; \hbox{finite} \; ,
\label{eq:46}
\ena
converges to the correct answer. Moreover, if the iteration is
truncated at $N=N_\mathrm{cut}$ and independently repeated many times,
the final values $a_k^{(N_\mathrm{cut})}$ are normal distributed with
the true value $a_k$ as mean. This paves the path to a bootstrap
analysis to obtain an error estimate for our estimate for $a_k$. 
A common choice is ($0 < \gamma \leq 1$) 
\be
\alpha _n \; = \; \left\{ \begin{array}{l c r}
  1 & \hbo \hbox{for} \hbo & 0 \le n \le n_t \; ,\\ 
  1/(n-n_t)^\gamma & \hbo \hbox{for} \hbo & n > n_t \; , 
\end{array} \right.
\label{eq:47}
\en
where the iterations with  $n \le n_t$ are considered as
thermalisation steps, and for which the limiting case $\gamma =1$ is
the optimal choice for error suppression. 

\vskip 1mm
\begin{figure*}[t]
\includegraphics[width=9cm]{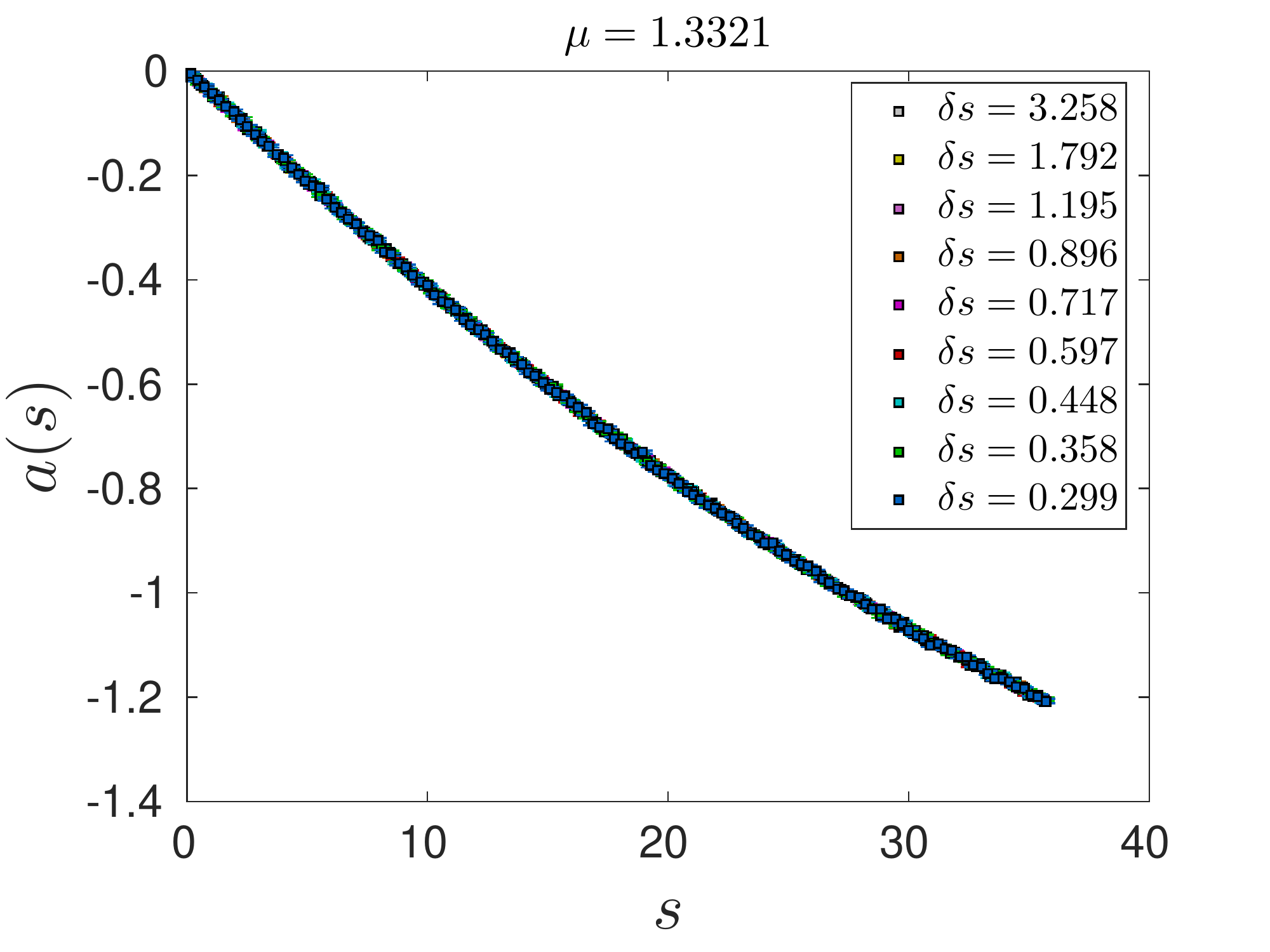} \hspace{0.2cm}
\includegraphics[width=9cm]{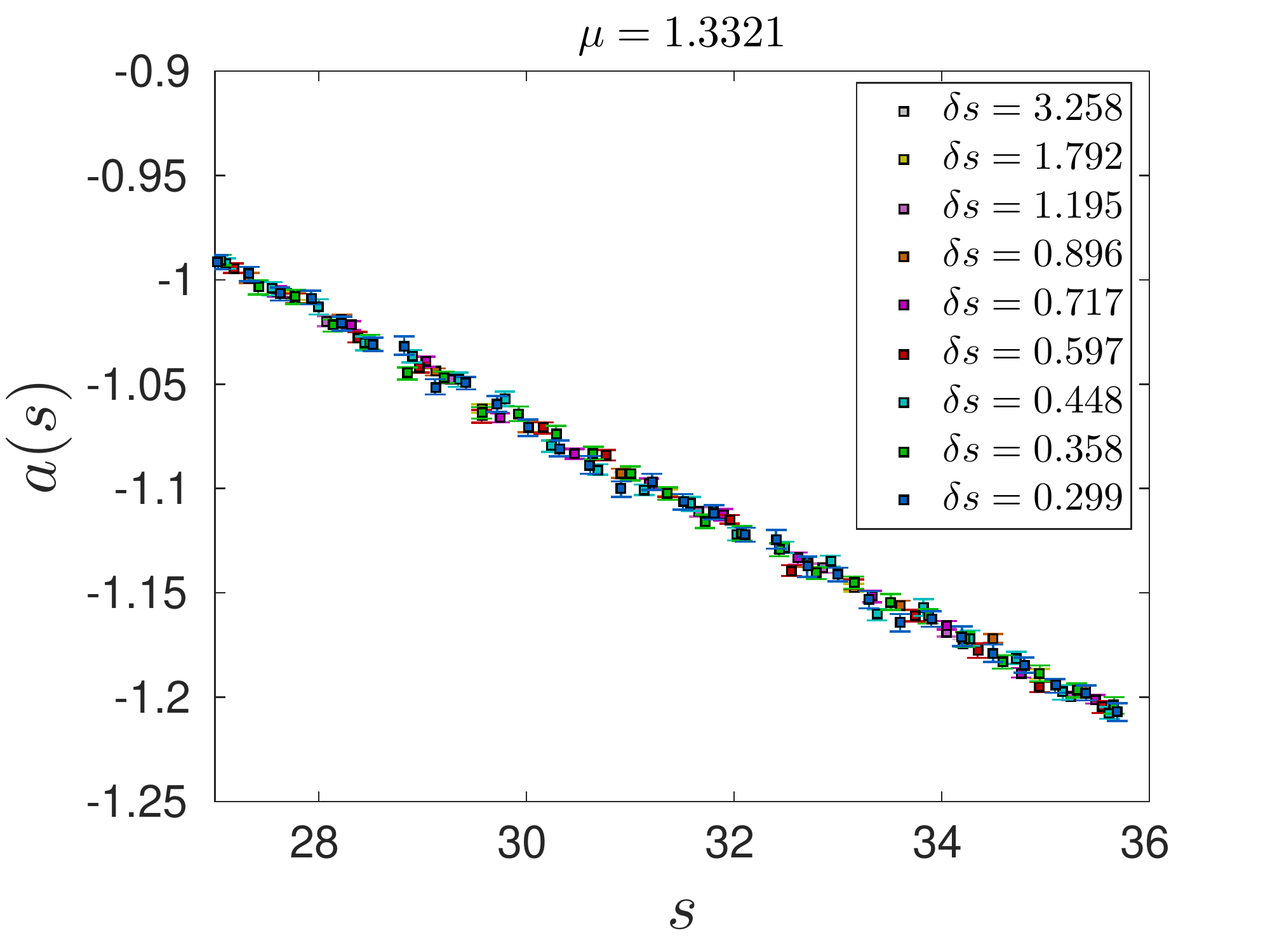} 
\caption{\label{fig:4} Strong Sign Problem regime: The LLR-coefficient
  $a(s)$ as a function 
  of $s$ for $\mu=1.3321$ (left panel).  Right: detail
  of the graph.  
}
\end{figure*}
Once the Taylor coefficients are obtained for the range $s$ of interest,
the generalised density-of-states $\rho (s)$ can be calculated in the
usual way:
\bea
\ln \, \rho (s)   &=&  -\sum_{k=1}^{n-1} a_i \, \ds  \; - \; a_n
\, \ds/2 \\  
n \; \hbox{such that:} && s_n \le s < s_{n+1} \; .
\label{eq:rhosum}
\ena 
Our final target is phase factor expectation value, which can be
obtained by means of two LLR integrals (details of the numerical
method will be presented in subsection~\ref{sec:LLR-int} below): 
\begin{equation}
\langle \e^{i\phi} \rangle=\frac{\displaystyle
\int_0^{s_{\rm max}} \rho(s) \cos(s) \; ds
}
{\displaystyle
\int_0^{s_{\rm max}} \rho(s) \; ds 
}
\label{eq:Lint}
\end{equation}
Since $\rho (s)$ is rapidly decreasing, we will find that it  is not
difficult to find a reliable cutoff $s_{\rm max}$.

\subsection{Thermalisation}

We find that the thermalisation is most demanding for small interval
sizes $\ds $ and for chemical potentials near the onset value.
In order to provide an insight into the thermalisation history, we
present here some results for the simulation parameters listed in
table~\ref{tab:1}. 

\begin{table}[t]
\begin{tabular}{cccccccc}
  $\ds $ & $s_k$ & $n_t$ & $\gamma$ & $L^4$ & $\beta $ & $\kappa $ &
  $\mu $ \\ \hline 
  $0.2986$ & $11.797$ & $30$ & $1$ & $8^4$ & $5.8$ & $0.12$ & $1.4321$  
\end{tabular}
\caption{\label{tab:1} Simulation parameters for one particular value
  $s$ }
\end{table}

\medskip
The thermalisation history for $40$ independent random starts is shown
in figure~\ref{fig:3}. Between each iterations, we performed $40$
sweeps at a fixed parameter $a_k^{(n)}$ in order to let the system
equilibrate.  

We see a decrease of the width of the error band with
increasing iteration number $n$, which is due to the Robbins Monro
underrelaxation. In the production runs for the results below, we have
chosen $n_t=200$ and a maximum of $1,000$ iterations. We then make use
of the Robbins  Monro feature that the final values for $a_k$ are
normal distributed with the correct mean. For the statistical
analysis, we repeated each iteration $40$ times and use the copies for
$a_k$ for the bootstrap analysis.

For a consistency check and to analyse the effect of the Robbins Monro
parameter $\gamma $, we calculated the average $a_k$ for different
values of $\gamma $. we find:

\medskip
\begin{tabular}{l|lllll}
  $\gamma $ & 0.6 & 0.7 & 0.8 & 0.9 & 1.0 \\ 
  $- a_k$ & 3.287 & 3.334 & 3.256 & 3.288 & 3.300 \\ 
  err $[10^{-2}]$ &  5.397 & 3.495 & 2.356 & 1.656 & 1.082
\end{tabular}

\medskip
We did not observe any ergodicity issues and found that the limiting
case $\gamma =1$ is most effective for error reduction as expected.

\subsection{Probability distribution of the imaginary part \label{sec:prob} }

According to figure~\ref{fig:1}, we will distinguish three parameter
regimes depending on the choice of the chemical potential $\mu $:
\begin{itemize}
\item[$\bullet$] {\it Low density regime} for $\mu \stackrel{<}{_\sim}
  1.1$: this regime 
  might be   accessible by a Taylor expansion with respect to $\mu $
  and simulations using reweighting.
\item[$\bullet$] Regime with a {\it strong sign problem} for $1.1
  \stackrel{<}{_\sim} \mu   \stackrel{<}{_\sim} 1.4$: this regime is
  beyond the scope of   standard Monte-Carlo methods and will be
  specifically targeted with   the LLR-method below. 
\item[$\bullet$] {\it Dense regime} for $1.4 \stackrel{<}{_\sim} \mu  \leq m\approx 1.427$:
    the system possesses a significant quark density, which reaches half
    of the saturation density for $\mu =m$. 
\end{itemize}
Because of the duality (\ref{eq:12}), we do not need to explicitly
explore the regime $\mu > m$. We stress that the above regime
boundaries have been chosen in an ad hoc way. We are not aware of any
physical phenomenon that would define these boundaries in a rigorous
way. The different regimes above, however, have quite distinct features
as we will reveal in this section by exploring the density-of-states.

\medskip
To this aim, we have calculated the LLR-coefficients $a_k$
(\ref{eq:40}) over a range of imaginary parts $s$ for given chemical
potentials. The simulation parameters again have been
$$
8^4 \hbo \beta=5.8 \hbo \kappa =0.12 \; . 
$$
Note that the LLR-method becomes exact in the limit of vanishing
interval size $\ds $.
In practice, we check that our result for $a_k$
does not dependent on $\ds $. 
We illustrate this fact for $\mu = 1.3321$, which belongs to the
interesting regime of a strong sign problem.
Our findings are shown in figure~\ref{fig:4}. We find that the
coefficients are quite insensitive to size of $\ds $. This also holds 
for the other regimes. Note that a smaller $\ds $ requests more
intervals to cover the same (integration) domain for $s$. We found
that $\ds = 0.896$ is a good compromise between accuracy and
computational effort, and it is this value which we have used in most
simulations.

\begin{figure}
\includegraphics[width=9cm]{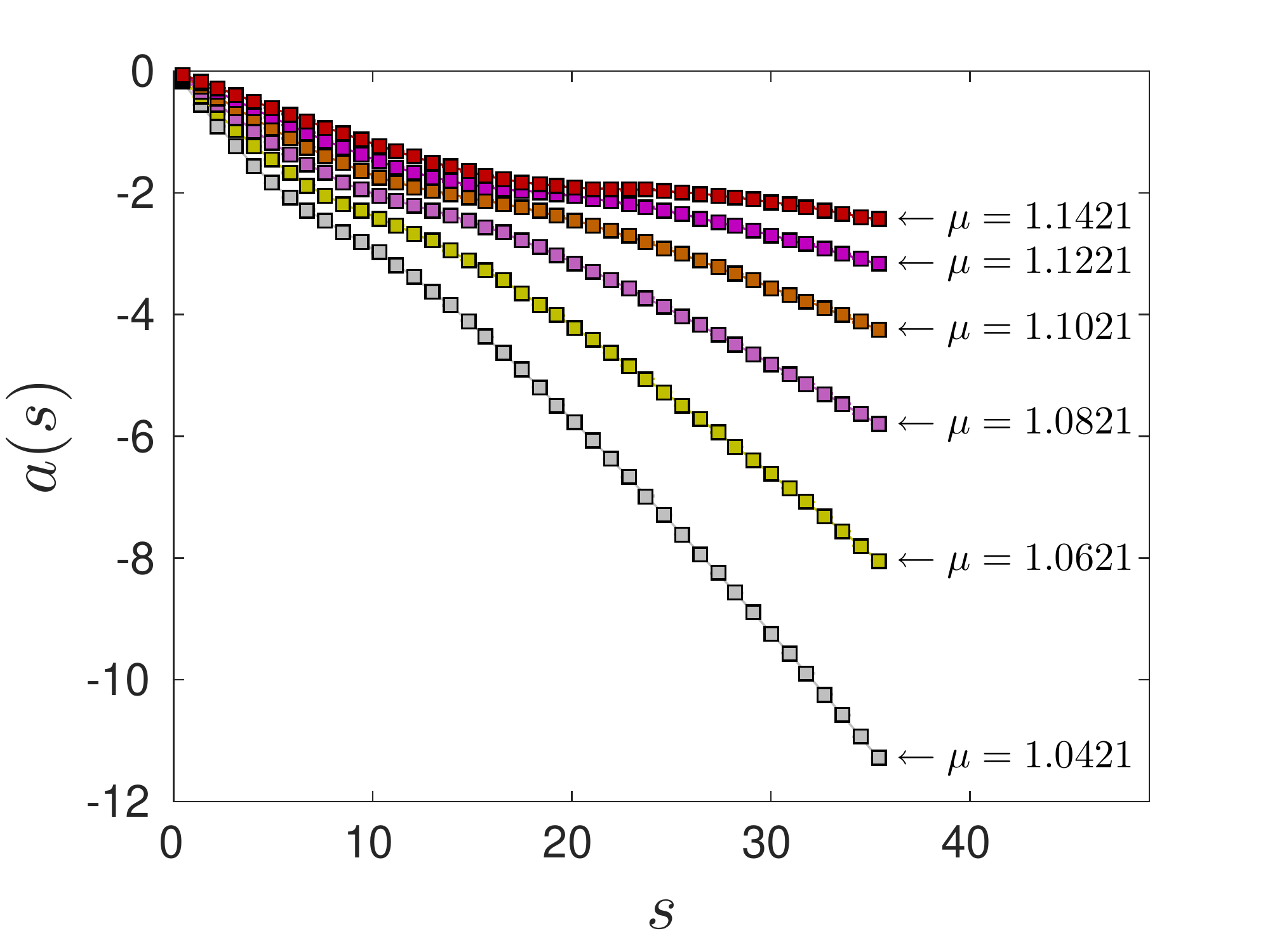} 
\caption{\label{fig:5} Low density regime:
  the LLR-coefficient $a(s)$
  as a function of $s$ for several values of the chemical potential $\mu$
  between $1.0421$ and $1.1421$. The error bars are smaller than the symbols.
  }
\end{figure}
\begin{figure}
\includegraphics[width=9cm]{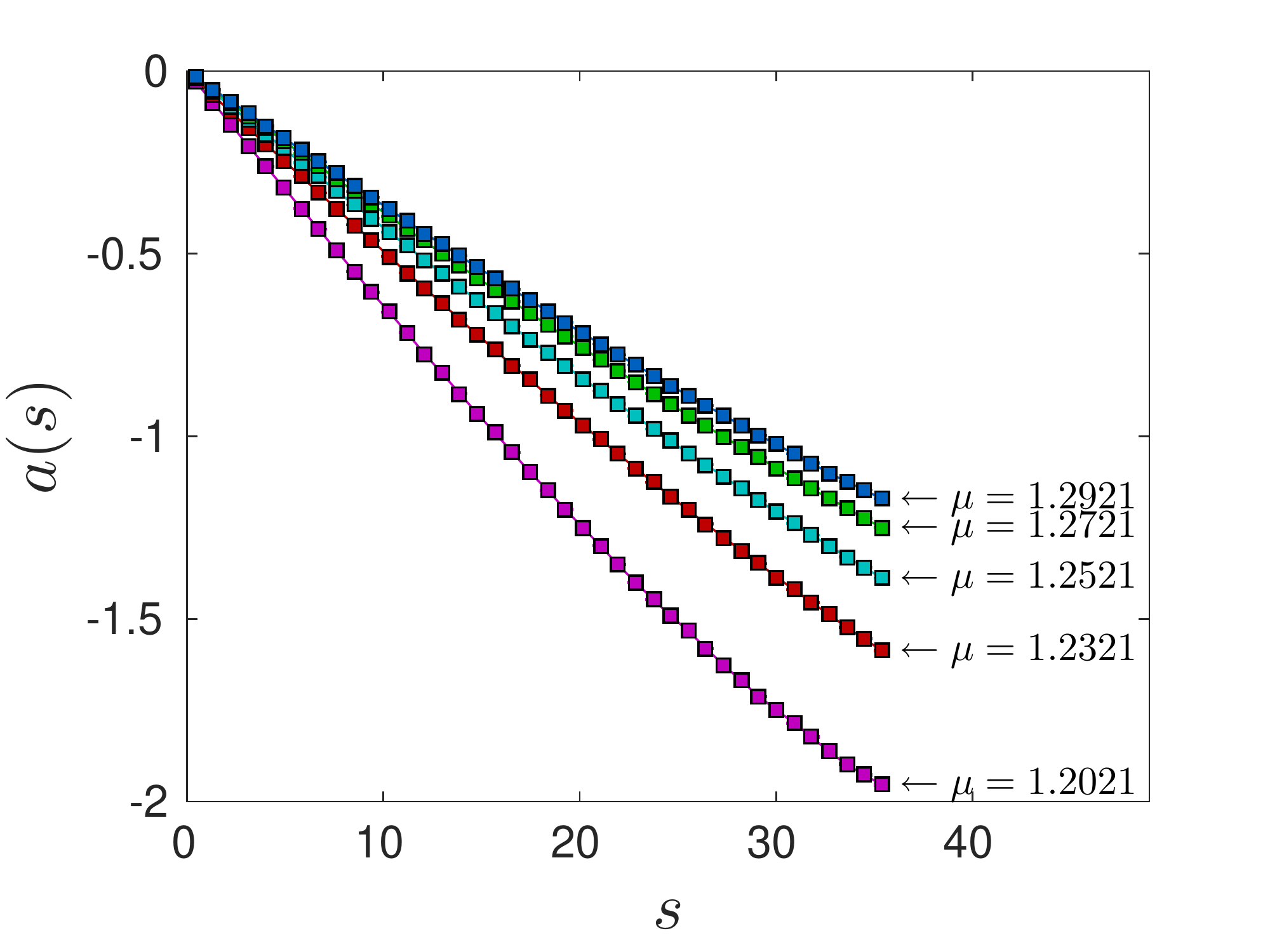} 
\caption{\label{fig:6} Strong sign problem regime (i): $a(s)$ for several
  values of the chemical potential $\mu$ between $1.2021$ and $1.2921$.
  Note that the y-scale differs from the previous plot. We observe that  
  $a(s)$ is an increasing function of $\mu$ for any value of $s>0$.
}
\end{figure}
\begin{figure}
\includegraphics[width=9cm]{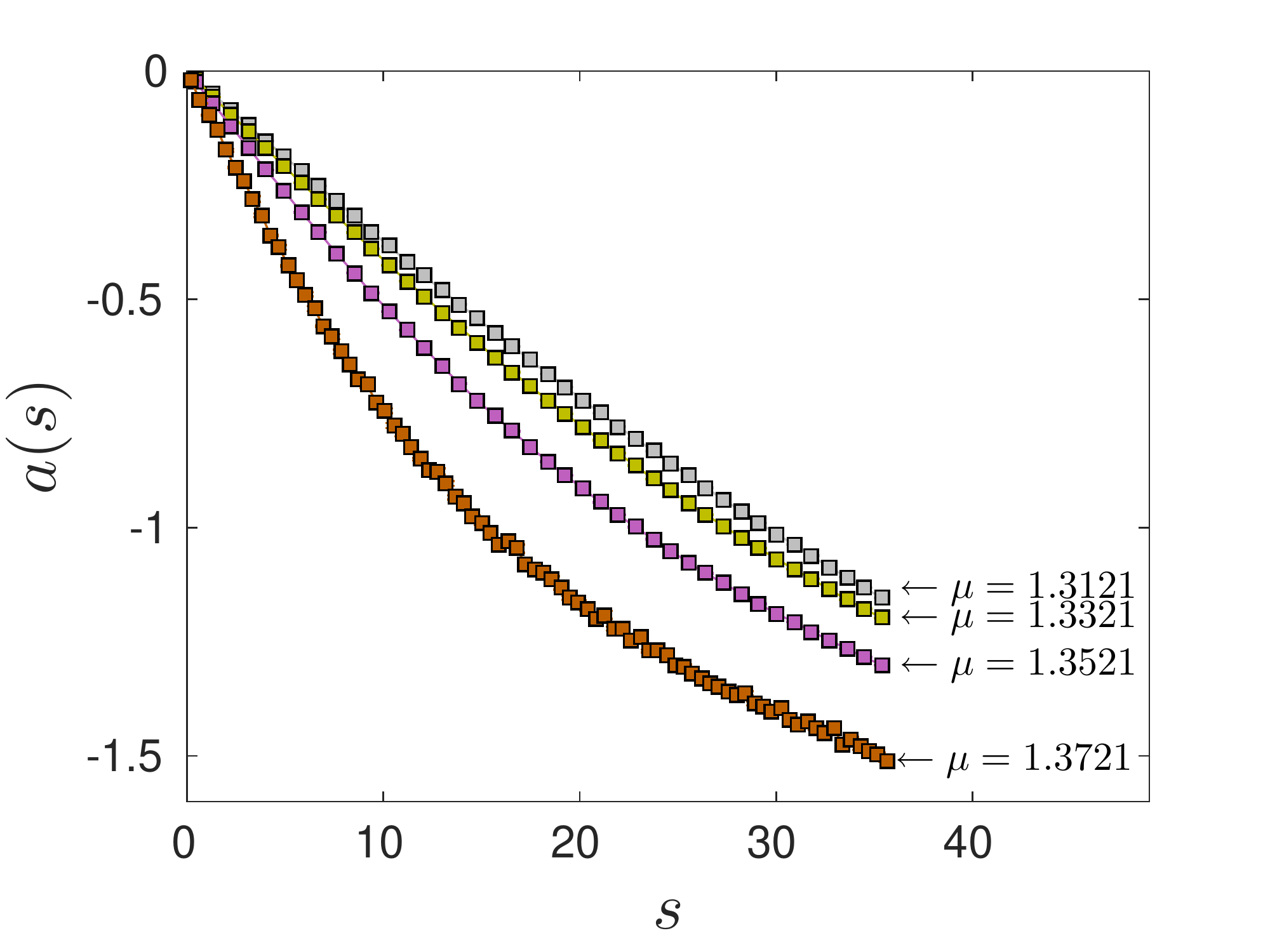} 
\caption{\label{fig:7} Strong sign problem regime (ii): $a(s)$ for several
  values of the chemical potential $\mu$  between $1.3121$ and $1.3721$.
  In this range, we observe that $a(s)$ is a decreasing function of $\mu$.
}
\end{figure}
\begin{figure*}
\includegraphics[width=9cm]{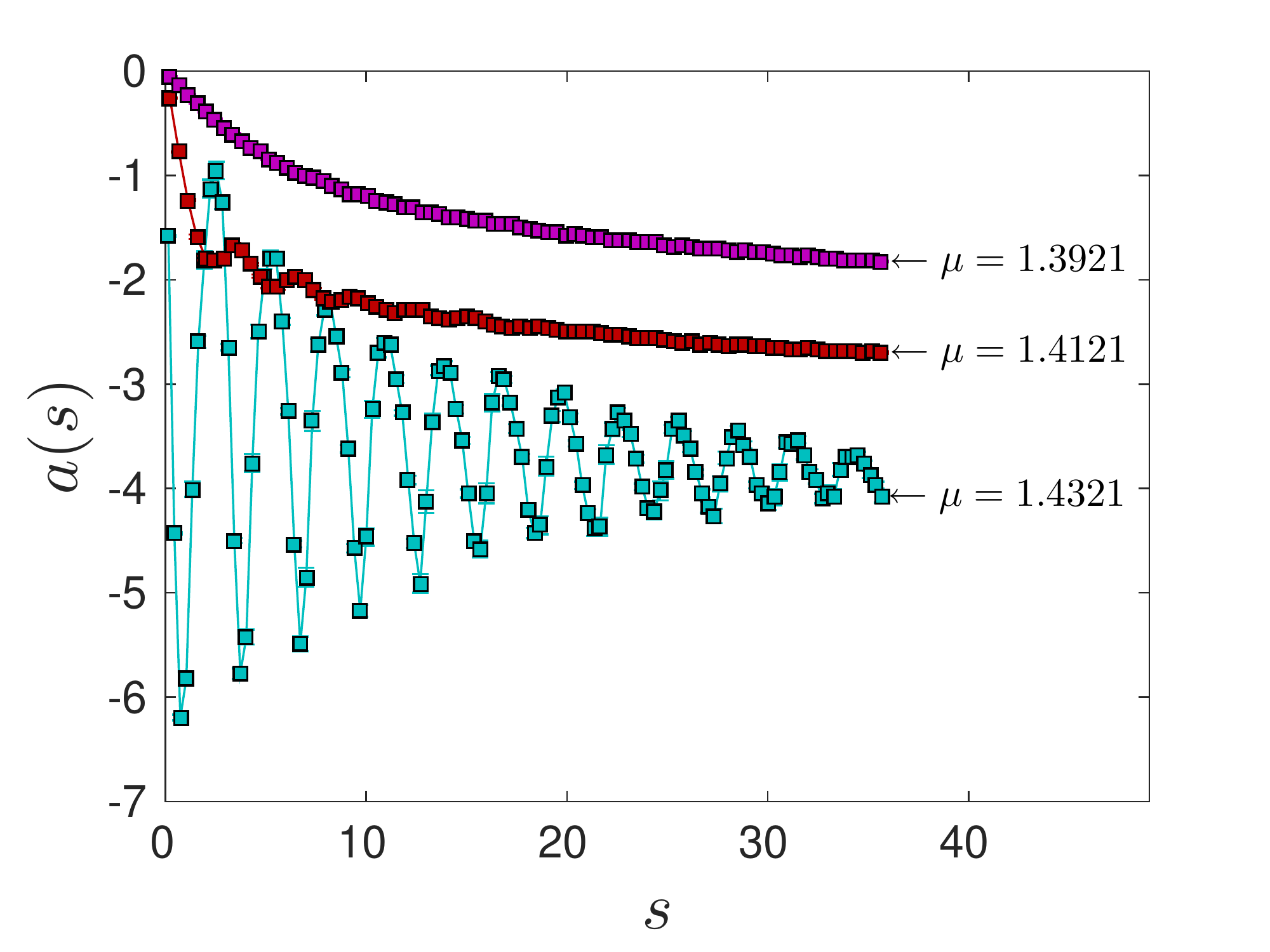} 
\includegraphics[width=9cm]{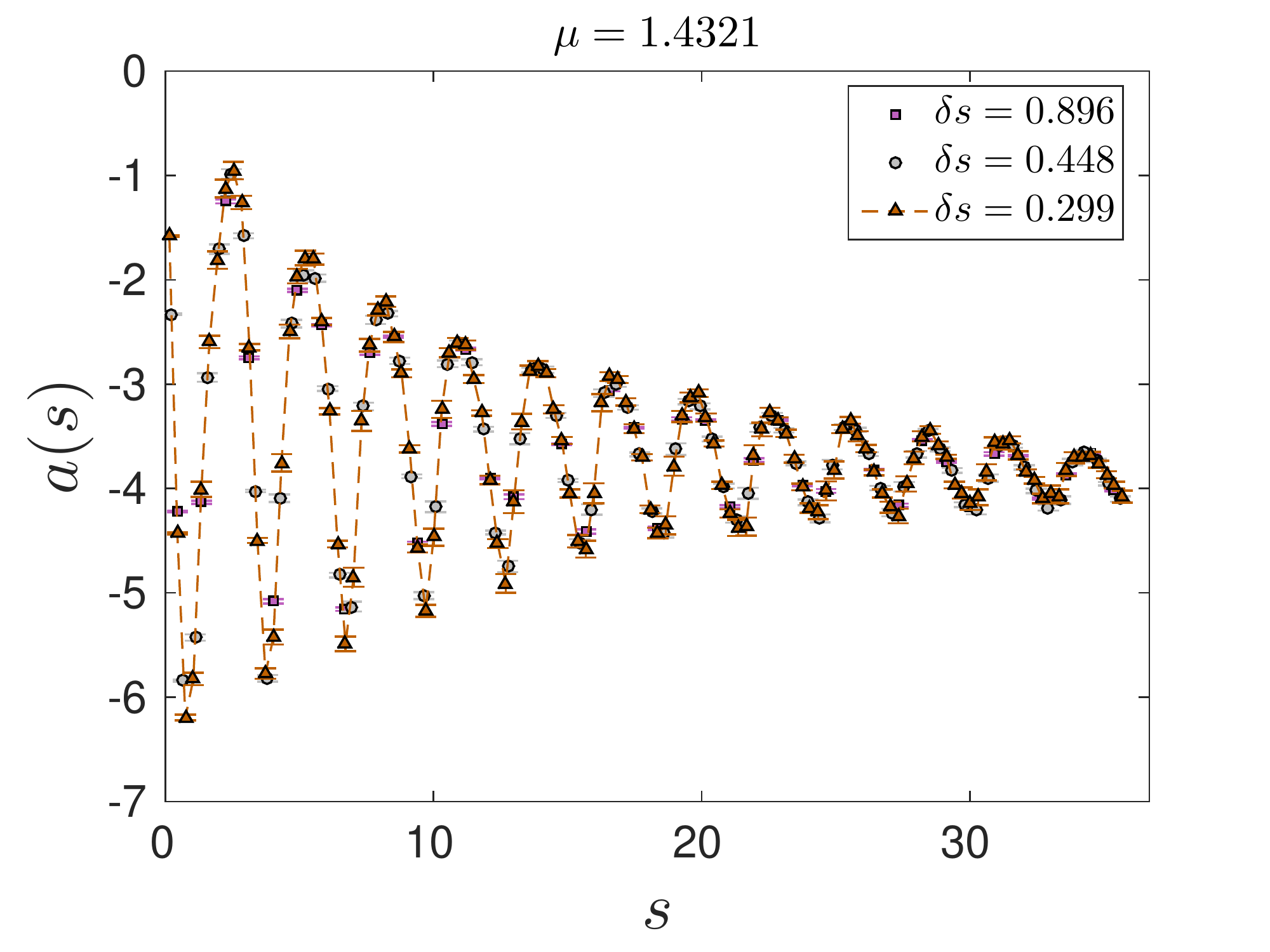} 
\caption{\label{fig:8} Left: Dense regime(i): $a(s)$ for several
    values of the chemical potential $\mu$ close to the mass threshold
  $m$. Right: Dense regime (ii): $a(s)$ as a function of $s$ near
  ``half-filling'' (slightly above $m$) for three different values of
  $\ds$. 
}
\end{figure*}
\begin{figure*}
  \includegraphics[width=9cm]{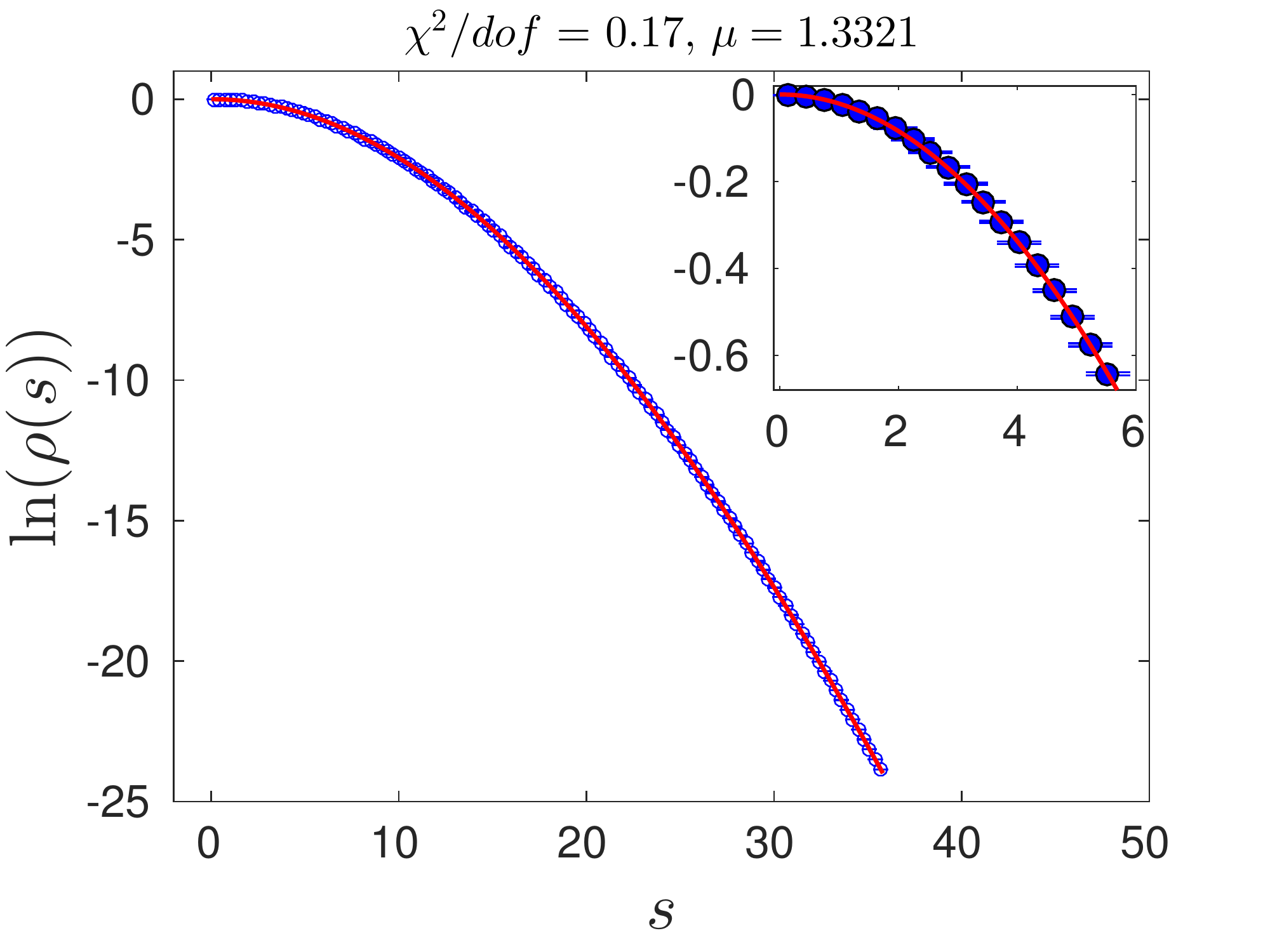}
  \includegraphics[width=9cm]{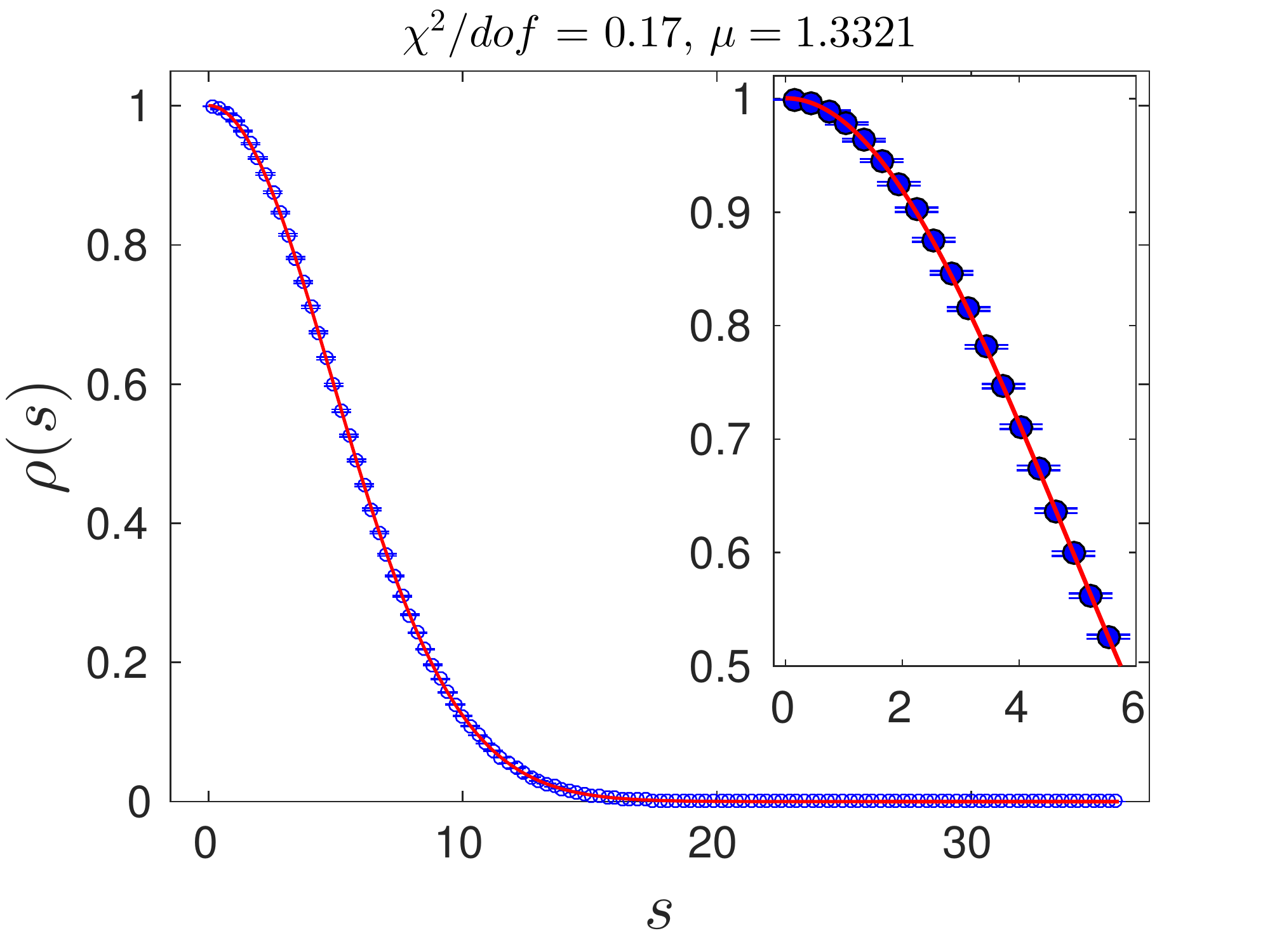}
\caption{\label{fig:10} Left: Natural logarithm of the density of
  states as a function   of $s$ for $\mu=1.3321$, in the strong sign
  problem regime.   We show both the data points with their error bars
  (blue cross) and   the fit results (red solid line) for $120$
  intervals   in $s$,   corresponding to $\delta_s\sim 0.299$. Right:
  Same as left, but on a linear scale. 
}
\end{figure*}

\medskip 
Figure~\ref{fig:5}-\ref{fig:8}, left panel, show the LLR-coefficient as
a function of $s$ for various values of the chemical potential.
We stress that in these figures, the error bars are present but smaller
than the symbols. Error bars are obtained from $40$ independent
sets of $a$ that are subjected to $500$ bootstrap samples.
Figure~\ref{fig:5} shows the low density regime. We find a slight
modulation of $a(s)$ with $s$, which did not occur for $\mu=1.3321$
(see figure~\ref{fig:4}). In figure~\ref{fig:6} and~\ref{fig:7}, we
summarise our findings 
for $a(s)$ for a range of chemical potentials that mostly belong to the
strong sign problem regime. We observe a quite distinct behaviour: the
curvature of the curves increases with increasing chemical
potential. For the largest values of $\mu $ shown in figure~\ref{fig:7}, we
enter the dense phase.
Our largest values of  $\mu $ are shown in~\ref{fig:8}, left panel. 
Here, we observe that $a(s)$ starts to show an
oscillatory behaviour. Needless to say that we have checked that these
oscillations are independent of the choice of $\ds $ and statistically
significant. This is illustrated in figure~\ref{fig:8}, left panel,
where we show the coefficient $a(s)$ for the chemical
potential $\mu=1.4321$, which is slightly above the mass threshold of 
$m=1.42711$.

\subsection{The LLR integration \label{sec:LLR-int}}

\begin{figure*} 
\includegraphics[width=9cm]{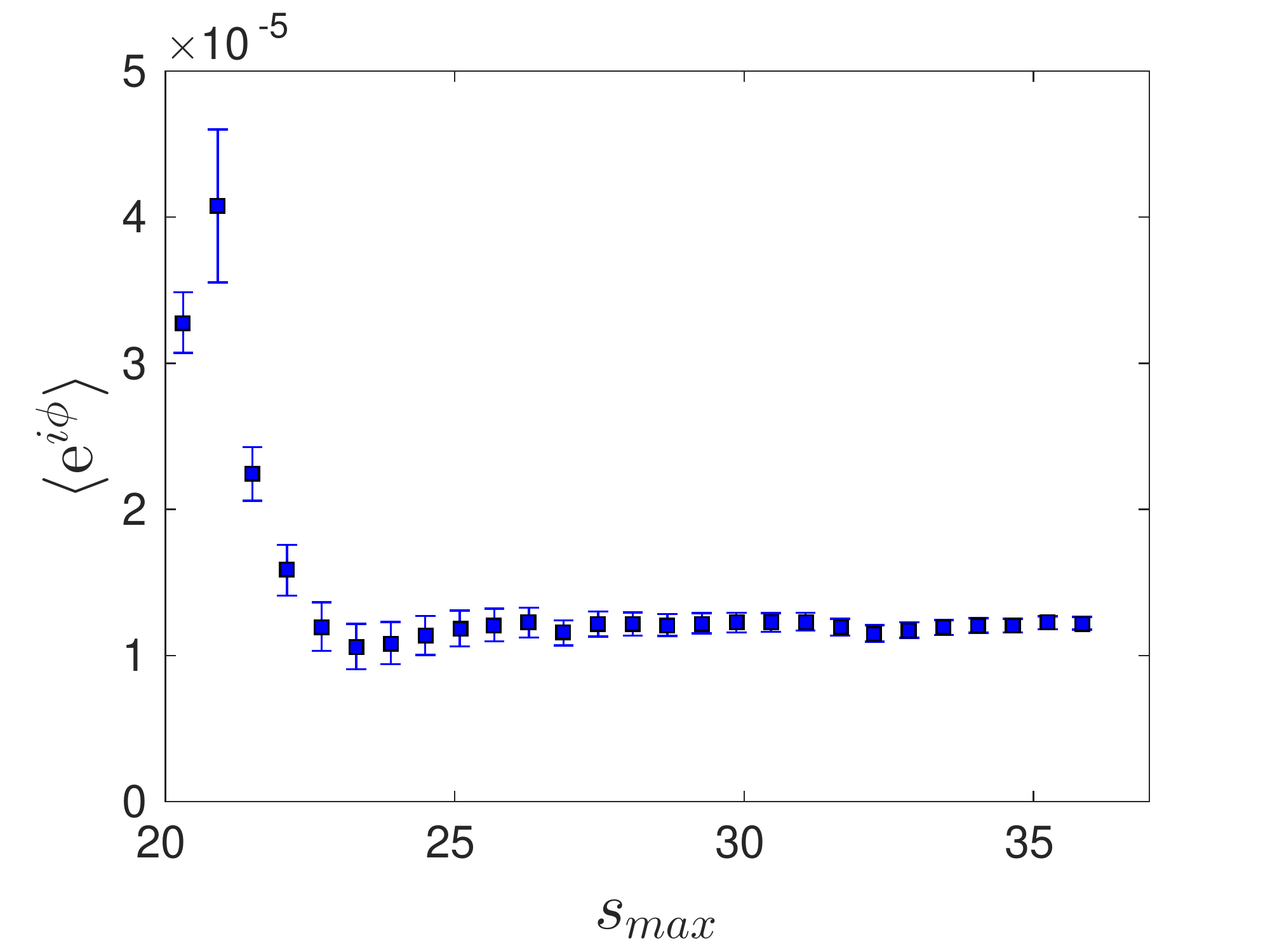}
\includegraphics[width=9cm]{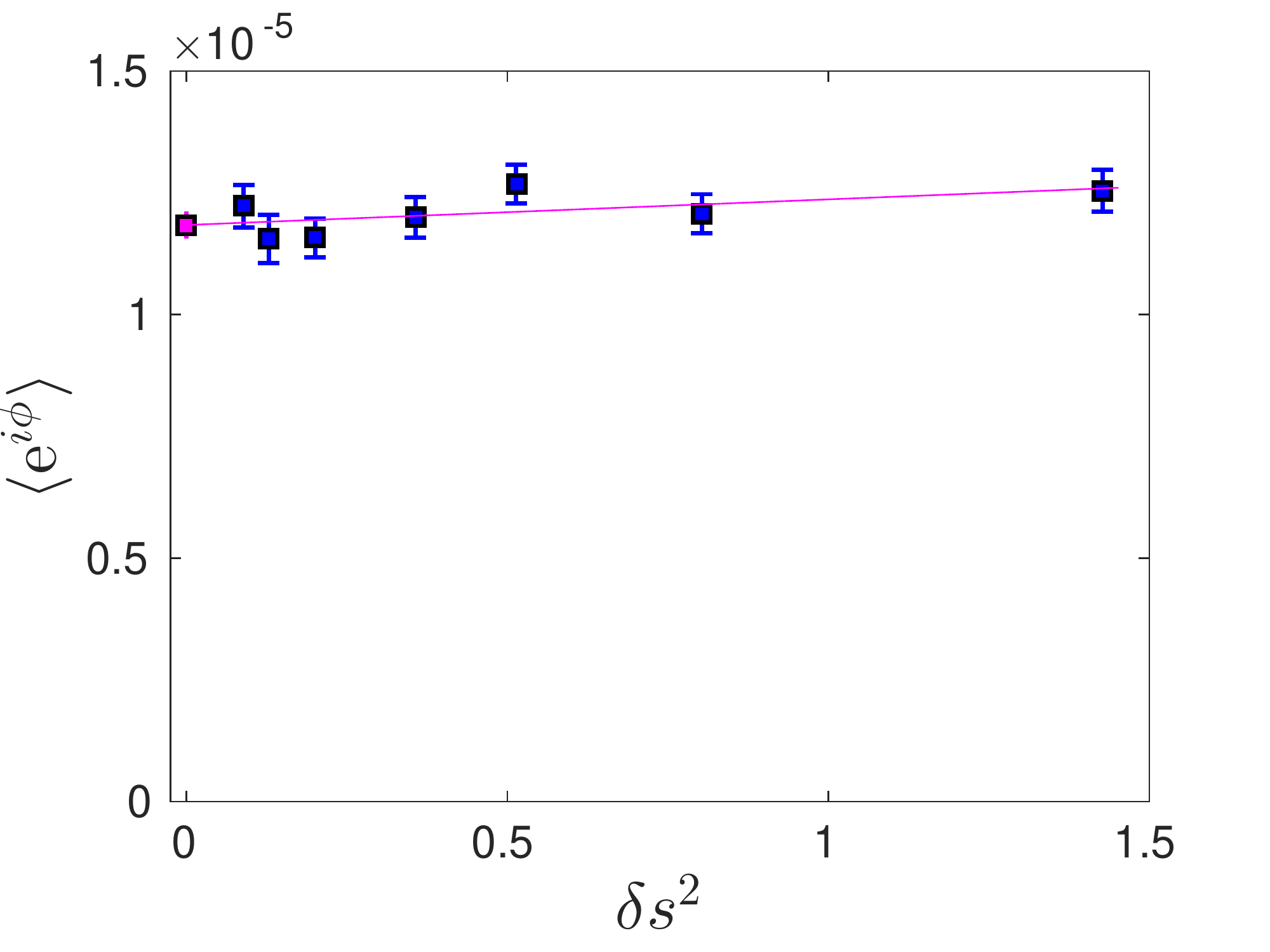}
\caption{\label{fig:11} Left: Result for the phase factor expectation
  value   as a function of the cut   $s_{\rm max}$. Results are shown
  for $\mu=1.3321$ and $\delta s = 0.29867$. Right:
  $\langle \e^{i\phi} \rangle$ for $\mu=1.3321$ as a function of
  $\ds^2$.   The results of the extrapolation is $1.185 (18) \times
  10^{-5}$,  statistical error only.
} 
\end{figure*}
Once the coefficient $a(s) $ have been extracted,
we are in a position to calculate the phase factor expectation value
$\langle {\rm e} ^{i \phi}\rangle $ for a given value of $\mu$
by means of (\ref{eq:Lint}). The straightforward method would be to
make use of the piece-wise linear interpolation (\ref{eq:rhosum}) and
to control the systematic errors in the Riemann sense by making $\ds $
smaller. It was already noted in~\cite{Langfeld:2014nta} for the case
of the $Z_3$ theory at finite densities that this method does not
muster enough precision at an affordable size $\ds $ to obtain a good
signal to noise ratio. Instead of seeking convergence in the Riemann
sense, we expand $\ln \rho $ in terms of basis functions $f_n(s)$:
\be
\ln \rho (s) \; = \; \sum _{n}^{N_\mathrm{max}} c_n \, f_n(s) \; .
\label{eq:50}
\en
The approximation now occurs by the truncation of the above sum at
$N_\mathrm{max}$. Here, we follow the strategy of {\it compressed
  sensing} (see e.g.~\cite{Candes2006}) and choose the basis
in such a way that a minimal number of coefficients $c_n$
represents the data at given accuracy and $\chi ^2$ per degree-of-freedom (dof) of the
fit. It is quite remarkable that a basis with simple powers of $s$,
i.e.,
\be
f_n(s) \; = \; s ^{n} \; .
\label{eq:52}
\en
already produces very good results, at least for the $Z_3$
theory~\cite{Langfeld:2014nta}. Eq.(\ref{eq:52}) is also our choice
here for HDQCD. Note that coefficients $c_n$, with $n$ are
incompatible with the theory's reflection symmetry $\ln \rho(-s) = \ln
\rho(s)$ and are therefore set to zero. 

\medskip
\parindent=0pt
In summary, our approach is: 
\begin{itemize}
\item[$\bullet$] 
  Using the numerical estimates $a_k$, we build the function
$P(s)=\ln(\rho(s))$ according to 
  \begin{eqnarray}
    \label{eq:P}
P(s)  &=& -\sum_{k=1}^{n-1} a_i \, \ds  \; - \; a_n \, \ds/2 \;, \\ 
s &=&  s_n + \ds /2 = n \ds + \ds /2  \;,
  \end{eqnarray}
  where in the last equation, we choose $s_0=0$ as a starting point.
\item We fit the result to a even-powers polynomial
\begin{equation}
P(s) =  \sum_{i=0}^{deg/2} {c_{2i}}\, s^{2i} \;.
\end{equation}
\item[$\bullet$] From the fit result, we reconstruct the density  
\begin{equation} 
\rho(s) = \exp(P(s))  \;.
\end{equation}
\item[$\bullet$] Finally, we semi-analytically compute the LLR integral 
\begin{equation}
\langle \e^{i\phi} \rangle=\frac{\displaystyle
\int_0^{s_{\rm max}} \rho(s) \cos(s) \; ds
}
{\displaystyle
\int_0^{s_{\rm max}} \rho(s) \; ds 
}
\end{equation}
\end{itemize} 
We have performed various checks in order to ensure that our procedure
is stable. First, we have tried different truncations: we denote by
$A_i$ a fit to a polynomial of degree $i$ in which all the coefficient
$c_{2i}$ are free parameters. We also performed some fits with $c_0$
fixed to $0$, we call them $\tilde A_i$. Some details of our fit
procedure for the finest $\delta s$  can be found in
Table~\ref{table:fitresults} for the specific value of $\mu=1.3321$. 
By comparing $\tilde A_2$ with $A_2$ and $A_6$ with $\tilde A_6$, 
we see that constant term $c_0$ has very little effect on the other
fit parameters.  All in all, we observe that the fit procedure is
robust, however our data are clearly best fitted by a degree-6
polynomial. Adding higher degrees gives compatible results with larger
errors (see $A_8$). We also present the fit results for $\ds=0.29867$
in Figure~\ref{fig:10}. 

Since we are looking for a very small signal emerging after large
cancellations, even the trivial identity 
\begin{equation}
\int_0^{s_{\rm max}}  \to \frac{1}{2} \int_{-s_{\rm max}}^{s_{\rm
    max}} \hbo \hbox{(folding)} 
\end{equation}
might perform differently upon its numerical implementation. In order to
check the robustness of our results, we implemented both integrals. 
In Table~\ref{table:fitresults},
the first integral (from $0$ to $s_{\rm max}$) is denoted by (i)
and the second (from $-s_{\rm max}$ to $s_{\rm max}$) is marked by (ii).
We see that the difference is smaller than the statistical error.

We have also checked that the results do not depend on the
cutoff $s_{\rm max}$, 
which is expected since $\rho (s)$ is rapidly decreasing.
This is illustrated in figure~\ref{fig:11}, left panel, where we have
changed the value of $s_{\rm max}$ before performing the fit of
$\ln(\rho)$, in other words we have varied the value of $n$ in the
functional form Eq.~\ref{eq:P}.  We have also checked that the
integral itself does not depend on  $s_{\rm max}$.

Finally we investigate the $\ds$ dependence. We have already seen
that the LLR coefficients exhibit very little dependence, but it remains
to be checked that the same holds for the LLR integrals leading to the
phase factor expectation value. In fact, we expect
the artefacts to be dominated by order $\delta s^2$ terms~\cite{Langfeld:2015fua}. Using $\mu=1.3321$ (from the severe 
sign problem region), we carried out simulations with several
different values of $\delta s$, reconstructed the LLR-coefficients and
finally performed the LLR-integrals to obtain values of $\langle \e
^{i \phi } \rangle$ for this set of $\ds $. We then performed a linear
extrapolation in  $\delta s^2$. Our findings are summarised in
figure~\ref{fig:11}, right panel: we indeed find a very small $\delta
s^2$ dependence. In fact, the final results for $\langle \e^{i\phi}
\rangle$ are more or less independent of $\ds $ within statistical
error bars. Our numerical findings for $\langle \e^{i\phi}
\rangle$ for different truncations can be found 
in Table~\ref{table:DPhi}.

\begin{table*}[tb]
\begin{tabular*}{\textwidth}{@{\extracolsep{\fill}}c|llllllll@{}}
\hline
$\ds =  0.29867$ 
& \multicolumn{1}{c}{$c_0 \times 10^{3}$}  & \multicolumn{1}{c}{$c_2 \times 10^{2}$} & \multicolumn{1}{c}{$c_4 \times 10^{6}$} 
& \multicolumn{1}{c}{$c_6 \times 10^{10}$} & \multicolumn{1}{c}{$c_8 \times 10^{14}$}  & \multicolumn{1}{c}{$\chi^2/dof$}
& \multicolumn{1}{c}{$\langle \e^{i\phi} \rangle\times 10^{5} \, (i)$} & \multicolumn{1}{c}{$\langle \e^{i\phi} \rangle\times 10^{5} \, (ii)$}  \\
\hline
$\tilde A_4$ &       &$-$2.0929(27) &   1.770(22)  &              &             &  7.4        &  0.962(16)  &  0.951(16)  \\
$A_4$ &$-$1.8(1.1)   &$-$2.0921(25) &   1.764(20)  &              &             &  7.3        &  0.957(15)  &  0.946(15)  \\            
$A_6$ &\phm0.3(1.0)  &$-$2.1148(44) &   2.423(84)  &$-$4.14(43)   &             &  0.14       &  1.222(44)  &  1.209(43)  \\
$\tilde A_6$ &       &$-$2.1145(47) &   2.418(91)  &$-$4.12(46)   &             &  0.15       &  1.220(47)  &  1.206(46)  \\
$\tilde A_8$ &       &$-$2.1161(68) &   2.507(270) &$-$5.46(2.71) & 6.03(11)    &  0.13       &  1.255(99)  &  1.241(98)  \\ 
\hline
\hline
\end{tabular*}
\caption{Fit results for $\mu=1.3321$ and $\delta s = 0.29867$. 
We show the fit coefficients for different truncations $A_i$, the
corresponding $\chi^2$ per degree of freedom and the result of the
integration. Missing results imply that the corresponding coefficient
is fixed to zero.  In the last rows, the results are obtained by
numerical integration either with or without folding. 
}
\label{table:fitresults}
\end{table*}

\begin{table*}
$$
\begin{tabular*}{\textwidth}{@{\extracolsep{\fill}}c|ll|ll|ll|ll@{}}
\hline
            & \multicolumn{2}{c|}{$A_4$}        & \multicolumn{2}{c|}{$A_6$}
            & \multicolumn{2}{c|}{$\tilde A_6$} & \multicolumn{2}{c}{$\tilde A_8$}  \\
\hline
$\ds$ 
& $\langle \e^{i\phi} \rangle\times 10^{5}$ & $\chi^2/dof$   & $\langle \e^{i\phi} \rangle\times 10^{5}$ & $\chi^2/dof$   
& $\langle \e^{i\phi} \rangle\times 10^{5}$ & $\chi^2/dof$   & $\langle \e^{i\phi} \rangle\times 10^{5}$ & $\chi^2/dof$   \\
\hline   
0.89600 & 0.944(17) & 6.1  &   1.196(39)  & 0.6   &  1.223(44) & 0.99 &  1.368(102) & 0.77 \\
0.71680 & 0.957(15) & 11   &  1.254(39)  & 0.8    & 1.299(43)  & 2.5  &  1.563(94) & 1.46 \\    
0.59733 & 0.929(14) & 6.1   &  1.189(41)   & 0.10  &  1.206(49) & 0.26 &  1.304(112) & 0.15  \\
0.44800 & 0.928(12) & 4.8    &  1.146(39)   & 0.16  &  1.151(46) & 0.18 &  1.159(112) & 0.18 \\
0.35840 & 0.923(16) & 3.7    &  1.144(49)   & 0.11  &  1.156(54) & 0.22 &  1.254(119) & 0.14 \\
0.29867 & 0.946(15) & 7.3   &   1.209(43)  & 0.14  &  1.206(46) & 0.15 &  1.241(98) &  0.13  \\ 
\hline
\end{tabular*}
$$
\caption{Result for the phase factor expectation value for  $\mu=1.3321$ as a function of $\ds$ 
for various fit Ans\"atze. 
The integral has been computed from $-s_{\rm max}$ to $s_{\rm max}$ 
(with folding).
}
\label{table:DPhi}
\end{table*}

\subsection{The phase factor expectation value} 

We have repeated the analysis outlined in the previous subsection for
several values of the chemical potential in the low density region,
in the strong sign problem and in the dense regimes (see
subsection~\ref{sec:prob} for a more formal definition of these
regimes). The numerical results are given in the Appendix. 
Each regime has its own challenges:

\medskip 
In the low density regime, the
LLR coefficients $a(s)$ are rapidly increasing with $s$. This implies a
rather narrow density-of-states $\rho (s)$, which might approximate a
Dirac function $\delta $ if $\mu $ approaches zero. Here, a careful
fine-tuning of $\ds $ and of the upper integration limit $s_\mathrm{max}$
would be in order. Since this regime is easily accessible by the
reweighting approach, we did not further pursue an optimal choice of
parameters, but used a generic choice of parameters for a validation
of the method only.

\medskip 
In the strong sign problem regime, our method works best: the results
are very robust against the parameter choice. The LLR coefficients
show a monotonic behaviour as a function of $s$, and the choice
of even powers of $s$ for the base functions $f_n(s)$ in (\ref{eq:52})
is converging rapidly: a few non-vanishing coefficients represents
hundreds of data points with a $\chi ^2 /$dof well below one.

\medskip 
The dense regime is obtained if the chemical potential takes values
close to the heavy quark mass, i.e., its onset value. The sign problem
in this regime is mild, and good results are obtained by the
reweighting approach. The coefficients $a(s)$ show oscillations around
a significant (negative) mean value. Upon reconstructing the
density-of-states (see (\ref{eq:41})), we find still find a monotonic
decreasing $\rho (s)$ (by virtue of the mean value of $a$), but
clearly a significant number of base functions $f_n(s)$ is needed to
grasp the oscillatory behaviour, and the method looses its
appeal. Insights into the cause of the oscillations would help to
develop a new set of base functions $f_n(s)$ that, again with few
coefficients, would grasp the essence of the numerical data. For the
present paper, we do present LLR results for this regime as well, but
observe that the representation of the data with the base functions
$f_n(s)=s^{2n}$ failed. Further work in this direction is needed,
which we will be presented elsewhere.

\medskip
Finally, we point out our rationale for the approximation of the
numerical data for $\ln \rho(s)$ in terms of $f_n(s)$: if few
base functions can approximate the data well ($\chi^2 / \mathrm{dof} <
1$), the bootstrap analysis for the final value of the phase factor
expectation values yields small statistical errors, and if the final
result is insensitive to the interval size $\ds $, we are confident
that the LLR approach solves the sign problem in this regime. We have
presented evidence for HDQCD in cases for which the reweighting method
can still produce statistical significant results. We also note 
that  if the base function fit fails in the sense that it produces a
$\chi^2 / \mathrm{dof} \ge 100$, it does not necessarily fail to
produce a result for the phase factor expectation value close to the
true value: it might that fit fails at a large scale in a region of
the integration parameter $s$ that is irrelevant to the final result
of the integration. We indeed have observed this for dense regime:
although the fit fails according to the obtained $\chi^2 /
\mathrm{dof}$, the final results is close to the value known from the
reweighting method.

\begin{figure} 
  \includegraphics[width=9cm]{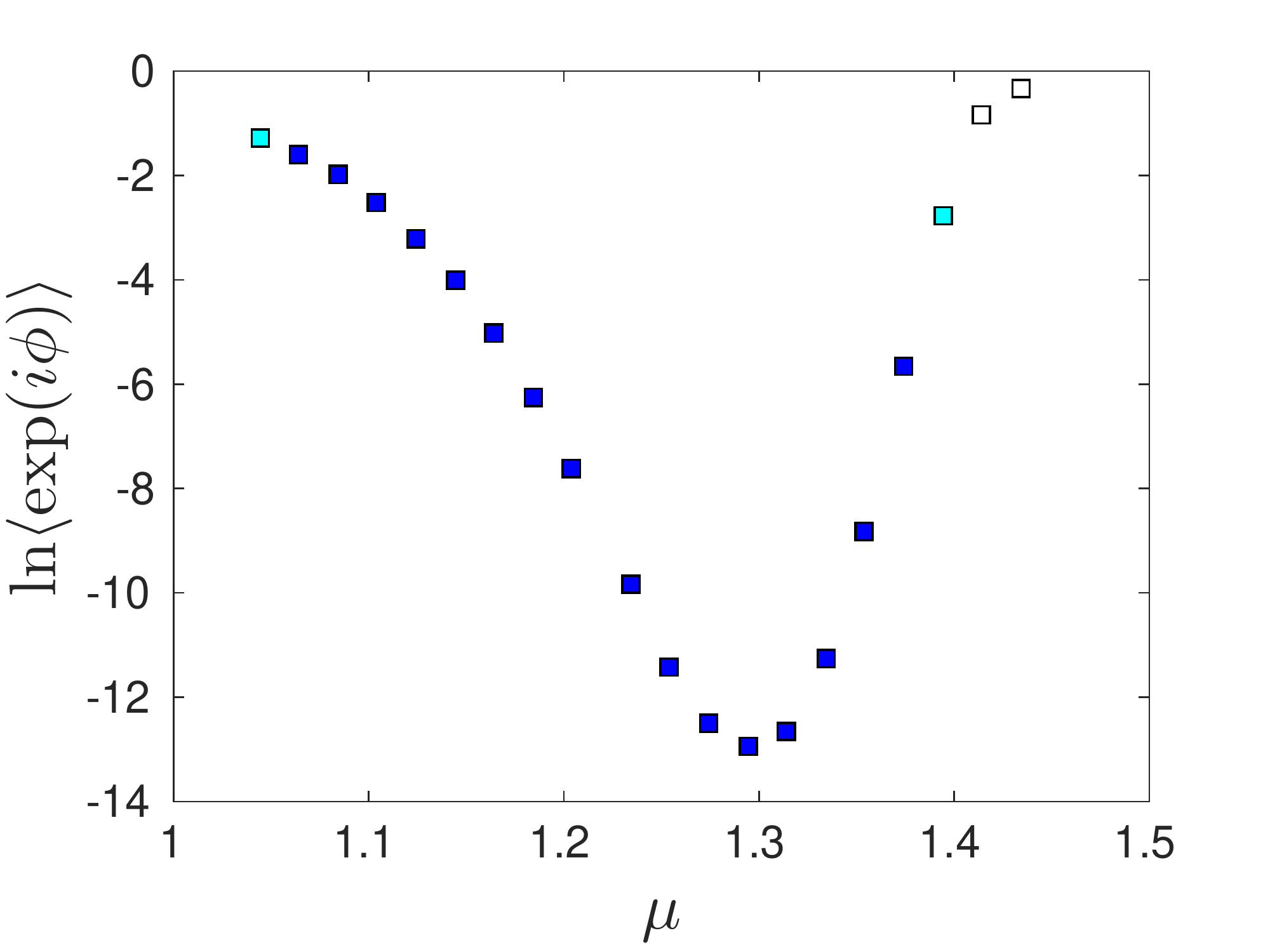}
  \caption{\label{fig:15} Natural logarithm of $\langle \e^{i\phi}
    \rangle$ for different values of $\mu$, 
    only the statistical errors are shown. The colour code is as
    follows: the plain blue points  (between $\mu=1.0621$
    and $\mu= 1.3721$) have a $\chi^2$ per degree-of-freedom of
    order one, the light blue points between $10$ and
    $50$, and the white points larger than $50$. 
  }
\end{figure}

\medskip
We finally present our main numerical finding. We are interested in
$\ln \, \langle \exp \{ i \phi \} \rangle $ since it is this quantity that
enters in e.g.~the calculation of the baryon density (see
(\ref{eq:15})): 
\be
\sigma (\mu) \; = \; \frac{T}{V} \, \frac{ \partial }{\partial \mu }
\, \ln \, \langle {\rm e}^{i\phi} \rangle +  \frac{T}{V} \, \frac{
 \partial }{\partial \mu } \ln \, Z_{PQ}(\mu) \; 
\label{eq:sigma}
\en
Our result for $\ln \, \langle \exp \{ i \phi \} \rangle $ as a
function of the chemical potential $\mu $ is shown in
figure~\ref{fig:15}. Further details, such as the quality of the fits
are given in the tables~\ref{table:fitallmu_low} - \ref{table:fitallmu}
in the Appendix.  We have also added these LLR results to the
figure~\ref{fig:1} of subsection~\ref{sec:rew} to validate the LLR
method against the reweighting data and to demonstrate the quality of
the LLR data in the strong sign problem regime. 

\section{Conclusions}

We have thoroughly studied QCD with a chemical potential for heavy
quarks using the density-of-states approach (LLR
version~\cite{Langfeld:2012ah,Langfeld:2014nta}). This approach 
allows for a determination of the probability distribution of the imaginary part of the
quark determinant featuring exponential error suppression. The
partition function appears as Fourier transform of this
probability distribution. We have bench-marked the LLR results against
results from the standard reweighting procedure (in the regime where
the latter produces a viable signal-to-noise ratio) and find excellent
agreement. We stress however that our approach yields an error that is
typically smaller by five orders of magnitude. 

\medskip
Due to an (approximate) particle hole duality at low temperatures, the
phase factor expectation value \break $\langle \exp \{ i \phi \} \rangle (\mu)$ is
symmetric around the onset chemical potential $\mu = m$ for
which $\langle \exp \{ i \phi \} \rangle = 1$. 
This suggests an {\it inverted Silver Blaze 
behaviour}: close to the mass threshold, the phase quenched baryon density 
{\it underestimates} the result of the full theory. 

\medskip
Depending on the chemical potential, we found three different regimes
which exhibit a different qualitative behaviour of the density-of-states
$\rho (s)$:

(i) In the low density regime, where the theory is almost
real, the domain of support of $\rho (s)$ is limited to small values
of $s$ as expected. 

(ii) For intermediate values of $\mu$, we find a strong sign problem
with $\langle \exp \{ i \phi \} \rangle (\mu) $ reaching values as
low as $10^{-6}$ for a small lattice size of $8^4$ (see~\ref{eq:31a}
for the simulation parameters).

(iii) For chemical potentials close to the onset value, the theory is
almost real again. By contrast to the low density regime, however, the
density-of-states for the imaginary part, i.e., $\rho (s)$, has a large
domain of support, and the corresponding LLR coefficients $a(s)$ show a  
oscillatory behaviour. It is exceedingly 
difficult to control the errors of the Fourier transform that is
needed to access the phase factor expectation value. Further studies
to explore the nature of the oscillations of $a(s)$ is left to
future work. We point out, however, the the regime close to onset is
accessible by reweighting.

\medskip
In summary, we find that the LLR approach to the probability
distribution of the imaginary part of the quark determinant is a
viable tool for the whole range of chemical potentials (with a
possible exemption near the onset transition). At least for the
moderate lattice size explored in this paper, the approach does solve
a strong sign problem.

\begin{acknowledgements}
We are grateful to B.~Lucini, A.~Rago and R.~ Pellegrini for helpful
discussions. We are grateful to the HPCC Plymouth for the support 
where the numerical computations have been carried out. NG and KL are
supported by  the  Leverhulme Trust (grant RPG-2014-118) and, KL by
STFC (grant ST/L000350/1). 
\end{acknowledgements}

\appendix
\section{Numerical details }

The tables below present details of the fit of the base function
expansion depending on the truncation (see section~\ref{sec:LLR-int}
for details). In boldface is the fit used for the final results
presented in figure~\ref{fig:15}. 

\input tables.tex


\bibliographystyle{spphys}       
\bibliography{density_ym}   

\end{document}

%% file: tables.tex
\begin{table*}
  \begin{tabular*}{\textwidth}{@{\extracolsep{\fill}}c|ll|ll|ll|ll|ll@{}}
\hline
$\mu\rightarrow$ &  \multicolumn{2}{c|}{$ 1.04210 $} &  \multicolumn{2}{c|}{$ 1.06210 $} &  \multicolumn{2}{c|}{$ 1.08210 $} &  \multicolumn{2}{c|}{$ 1.10210 $} &  \multicolumn{2}{c}{$ 1.12210 $} \\
\hline
&
$\langle \e^{i\phi} \rangle\times10$ & $\chi^2/dof$ & $\langle \e^{i\phi} \rangle\times10$ & $\chi^2/dof$ &
$\langle \e^{i\phi} \rangle\times10$ & $\chi^2/dof$ & $\langle \e^{i\phi} \rangle\times10^{2}$ & $\chi^2/dof$ &
$\langle \e^{i\phi} \rangle\times10^{4}$ & $\chi^2/dof$ \\
\hline
 $\tilde A_6    $&$ 2.216(4)$ & $2074$ & $1.577(4)$ & $1156$ & $1.126(3)$ & $1258  $ & $7.16(3)$ & $489 $ & $4.30(1)$ & $94$  \\
 $A_6           $&$ 2.160(4)$ & $1636$ & $1.548(4)$ & $1032$ & $1.104(3)$ & $1087  $ & $7.06(3)$ & $419 $ & $4.27(1)$ & $70$  \\
 $\tilde A_8    $&$ 2.513(5)$ & $786 $ & $1.834(6)$ & $293$  & $1.299(3)$ & $189   $ & $8.14(3)$ & $53  $ & $4.46(2)$ & $64$  \\
 $A_8           $&$ 2.452(5)$ & $610 $ & $1.806(5)$ & $24$   & $1.278(3)$ & $126   $ & $8.03(3)$ & $17  $ & $4.40(2)$ & $47$  \\
 $\tilde A_{10} $&$ 2.727(5)$ & $252 $ & $1.990(7)$ & $66$   & $1.372(4)$ & $45    $ & $8.32(4)$ & $43  $ & $4.29(2)$ & $43$  \\
 $A_{10}        $&$ 2.670(6)$ & $181 $ & $1.963(6)$ & $42$   & $1.349(4)$ & $8.3   $ & $8.18(3)$ & $10  $ & $4.20(2)$ & $16$  \\
 $\tilde A_{12} $&$ 2.856(6)$ & $75  $ & $2.072(7)$ & $20$   & $1.391(4)$ & $38.2  $ & $8.24(4)$ & $42  $ & $4.17(3)$ & $35$  \\
 $A_{12}        $&$\bf 2.806(6)$ & $42  $ & $\bf 2.042(7)$ & $4.8$  & $\bf 1.362(4)$ & $5.1   $ & $\bf 8.05(4)$ & $5.9 $ & $\bf 4.03(2)$ & $1.25 $  \\
 $\tilde A_{14} $&$ 2.927(7)$ & $41  $ & $2.107(9)$ & $40$   & $1.390(5)$ & $47    $ & $8.15(4)$ & $58  $ & $4.17(3)$ & $37$  \\
 $A_{14}        $&$ 2.878(6)$ & $67  $ & $2.072(8)$ & $31$   & $1.353(5)$ & $13    $ & $7.89(4)$ & $20  $ & $3.99(2)$ & $19$  \\
\hline   
\end{tabular*}
  \caption{Fit result of the the phase factor expectation value for the low values of $\mu$
    and different truncations. 
}
\label{table:fitallmu_low}
\end{table*}
\begin{table*}
  \begin{tabular*}{\textwidth}{@{\extracolsep{\fill}}c|ll|ll|ll|ll|ll@{}}
    \hline
$\mu\rightarrow$ &  \multicolumn{2}{c|}{$ 1.14210 $} &  \multicolumn{2}{c|}{$ 1.16210 $} &  \multicolumn{2}{c|}{$ 1.18210 $} &  \multicolumn{2}{c|}{$ 1.20210 $} &  \multicolumn{2}{c}{$ 1.23210 $} \\
\hline
&
$\langle \e^{i\phi} \rangle\times10^2$ & $\chi^2/dof$ & $\langle \e^{i\phi} \rangle\times10^3$ & $\chi^2/dof$ &
$\langle \e^{i\phi} \rangle\times10^3$ & $\chi^2/dof$ & $\langle \e^{i\phi} \rangle\times10^{4}$ & $\chi^2/dof$ &
$\langle \e^{i\phi} \rangle\times10^{5}$ & $\chi^2/dof$ \\
\hline
 $\tilde A_6    $ & 2.018(5)  & 117  & 6.817(47) & 41   & 1.888(5) & 12   & 4.89(4)  & 17    & 5.02(4)  & 13   \\
 $A_6           $ & 2.008(4)  & 115  & 6.710(24) & 32   & 1.883(5) & 11   & 4.78(3)  & 0.3   & 4.87(3)  & 0.3  \\
 $\tilde A_8    $ & 1.861(9)  & 37   & 6.415(43) & 21   & 1.959(1) & 0.7  & 5.02(6)  & 16    & 5.28(10) & 12   \\
 $A_8           $ & 1.832(8)  & 22   & 6.250(36) & 4.6  & 1.953(9) & 0.2  & 4.81(4)  & 0.3   & 4.94(8)  & 0.2  \\
 $\tilde A_{10} $ & 1.777(12) & 25   & 6.752(61) & 14   & 1.960(2) & 0.7  & 5.26(9)  & 16    & 5.86(21) & 12   \\
 $A_{10}        $ & 1.725(11) & 2.4  & 6.515(50) & 0.5  & 1.951(1) & 0.2  & 4.86(8)  & 0.2   & 5.11(17) & 0.1  \\
 $\tilde A_{12} $ & 1.817(16  & 234  & 6.930(80) & 13   & 1.945(2) & 0.6  & 5.67(16) & 15    & 6.99(46) & 11   \\
 $A_{12}        $ & 1.747(14) & 1.9  & \bf 6.605(53) & 0.2  & \bf 1.929(2) & 0.1  & \bf 4.98(13) & 0.2   & \bf 5.43(36) & 0.1  \\
 $\tilde A_{14} $ & 1.885(21) & 20   & 6.972(11) & 14   & 1.958(3) & 0.6  & 6.23(25) & 50    & 8.79(89) & 19   \\
 $A_{14}        $ & \bf 1.796(22) & 0.3  & 6.505(82) & 0.1  & 1.933(3) & 0.1  & 5.10(22) & 40.42 & 5.74(73) & 0.5  \\
\hline   
\end{tabular*}
  \caption{Fit result of the the phase factor expectation value for the middle-low values of $\mu$
    and different truncations. 
}
\label{table:fitallmu_med}
\end{table*}

\begin{table*}
  \begin{tabular*}{\textwidth}{@{\extracolsep{\fill}}c|ll|ll|ll|ll|ll@{}}
\hline
$\mu\rightarrow$
&  \multicolumn{2}{c|}{$ 1.25210 $} &  \multicolumn{2}{c|}{$ 1.27210 $} &  \multicolumn{2}{c|}{$ 1.29210 $} &  \multicolumn{2}{c|}{$ 1.31210 $} &  \multicolumn{2}{c}{$ 1.33210 $} \\
\hline 
&
$\langle \e^{i\phi} \rangle\times10^5$ & $\chi^2/dof$ & $\langle \e^{i\phi} \rangle\times10^6$ & $\chi^2/dof$ &
$\langle \e^{i\phi} \rangle\times10^6$ & $\chi^2/dof$ & $\langle \e^{i\phi} \rangle\times10^{6}$ & $\chi^2/dof$ &
$\langle \e^{i\phi} \rangle\times10^{5}$ & $\chi^2/dof$ \\
\hline
 $\tilde A_6    $ & 1.20(2)  & 12     & 3.97(6)   & 14   & 2.56(3)  & 9.8  & 3.33(4)   & 9.3   & 1.25(10) & 10.5  \\
 $A_6           $ & 1.16(1)  & 0.2    & 3.76(5)   & 0.3  & 2.47(3)  & 0.2  & 3.23(3)   & 0.2   & 1.21(9) & 0.9    \\
 $\tilde A_8    $ & 1.26(3)  & 12     & 4.26(13)  & 14   & 2.77(9)  & 9.5  & 3.60(12)  & 8.7   & 1.37(3) & 8.1    \\
 $A_8           $ & 1.15(3)  & 0.1    & 3.74(10)  & 0.4  & 2.51(7)  & 0.2  & 3.30(10)  & 0.1   & 1.28(2) & 0.1    \\
 $\tilde A_{10} $ & 1.35(7)  & 12     & 5.15(3)   & 14   & 3.09(27) & 9.7  & 4.04(28)  & 8.7   & 1.49(6) & 7.9    \\
 $A_{10}        $ &\bf 1.09(6)  & 0.1    &\bf 3.75(2)   & 0.4  &\bf 2.37(21) & 0.1  &\bf 3.23(21)  & 0.1   &\bf 1.27(5) & 0.2    \\
 $\tilde A_{12} $ & 1.60(2)  & 12     & 7.56(9)   & 13   & 3.97(63) & 9.8  & 5.03(64)  & 8.7   & 1.71(11) & 7.9    \\
 $A_{12}        $ & 1.02(14) & 0.1    & 3.98(7)   & 0.4  & 2.07(5)  & 0.1  & 3.01(47)  & 0.1   & 1.23(9) & 0.1    \\
 $\tilde A_{14} $ & 2.11(33) & 1392   & 1.34(21)  & 34   & 5.93(15) & 394  & 7.68(141) & 184   & 2.35(23) & 7.8  \\
 $A_{14}        $ & 0.829(28) & 2853  & 0.50(174) & 21   & 9.86(11) & 462  & 2.91(96)  & 246   & 1.34(17) & 0.4
\end{tabular*}
  \caption{Fit result of the the phase factor expectation value for the middle-high values of $\mu$
    and different truncations.  
}
\label{table:fitallmu_dense}
\end{table*}

\begin{table*}
  \begin{tabular*}{\textwidth}{@{\extracolsep{\fill}}c|ll|ll|ll|ll|ll@{}}
\hline
$\mu\rightarrow$
 &  \multicolumn{2}{c|}{$ 1.35210 $} &  \multicolumn{2}{c|}{$ 1.37210 $} &  \multicolumn{2}{c|}{$ 1.39210 $} &  \multicolumn{2}{c|}{$ 1.41210 $} &  \multicolumn{2}{c}{$ 1.43210 $} \\
\hline 
&
$\langle \e^{i\phi} \rangle\times10^4$ & $\chi^2/dof$ & $\langle \e^{i\phi} \rangle\times10^3$ & $\chi^2/dof$ &
$\langle \e^{i\phi} \rangle\times10^2$ & $\chi^2/dof$ & $\langle \e^{i\phi} \rangle\times10$ & $\chi^2/dof$ &
$\langle \e^{i\phi} \rangle\times10$ & $\chi^2/dof$ \\
\hline
 $\tilde A_6    $ & 1.354(7)  & 44   & 2.66(2)  & 47   & 3.42(2) & 858 & 2.177(6) & 4418  & 4.667(7) & \gchi  \\
 $A_6           $ & 1.316(6)  & 17   & 2.65(2)  & 43   & 3.26(2) & 735 & 2.135(6) & 3906  & 4.600(7) & \gchi   \\
 $\tilde A_8    $ & 1.523(12) & 17   & 3.24(4)  & 8.1  & 4.80(3) & 243 & 2.869(9) & 2478  & 5.564(6) & \gchi  \\
 $A_8           $ & 1.456(9)  & 0.3  & 3.22(4)  & 4.8  & 4.62(3) & 208 & 2.812(9) & 2107  & 5.484(7) & \gchi   \\
 $\tilde A_{10} $ & 1.613(25) & 16   & 3.60(7)  & 3.6  & 5.82(5) & 74  & 3.479(1) & 1574  & 6.207(6) & \gchi  \\
 $A_{10}        $ &\bf 1.489(19) & 0.1  &\bf 3.56(7)  & 0.7  & 5.65(4) & 63  & 3.404(1) & 1296  & 6.117(7) & \gchi  \\
 $\tilde A_{12} $ & 1.717(47) & 15   & 3.80(12) & 3.1  & 6.51(6) & 26  & 3.994(1) & 1065  & 6.659(6) & \gchi   \\
 $A_{12}        $ & 1.486(36) & 0.1  & 3.73(12) & 0.3  &\bf 6.35(5) & 21  & 3.902(1) & 855   & 6.560(6) & \gchi   \\
 $\tilde A_{14} $ & 1.926(77) & 63   & 3.96(16) & 4.5  & 6.97(7) & 21  & 4.422(1) & 944   & 7.021(5) & \gchi   \\
 $A_{14}        $ & 1.511(58) & 58   & 3.85(16) & 1.8  & 6.83(5) & 21  &\bf 4.317(1) & 799   & 6.915(6) & 4858  \\
 $A_{16}        $ & -         & -    & 3.95(22) & \gchi & 7.17(6) & \gchi & 4.658(15) & \gchi  & \bf 7.224(6) & 4787  \\
\hline
\end{tabular*}
  \caption{Fit result of the the phase factor expectation value for the high values of $\mu$
    and different truncations. We do not give the $\chi^2$-values larger than $5000$. 
}
\label{table:fitallmu}
\end{table*}